\documentclass[11pt,a4paper]{article}

\usepackage[utf8]{inputenc}
\usepackage{mathrsfs,amsmath} 
\usepackage{systeme}
\usepackage{mathtools}
\usepackage{authoraftertitle}
\usepackage{physics}
\usepackage{hyperref}
\usepackage{subcaption} 
\usepackage{lipsum}
\usepackage[draft]{todonotes} 
\usepackage{chngcntr}
\usepackage{tikz-cd}
\usepackage{chemfig}
\usepackage{needspace}

\usepackage[absolute]{textpos}

\usepackage{mathrsfs,amsmath} 
\usepackage[section]{placeins}
\usepackage{subfiles}

\usepackage[math]{cellspace}

\usepackage{graphicx,epsf}
\usepackage{array,multirow}
\usepackage{bm}

\renewcommand{\vec}[1]{\boldsymbol{#1}}
\newcommand{\uvec}[1]{\boldsymbol{\hat{\textbf{#1}}}}

\newcommand{\ySubFig}[1]{
	\begin{subfigure}[t]
		{0.49\textwidth}
		\includegraphics[width=\textwidth]{./#1}
	\end{subfigure}
}

\newcommand{\yFigTwo}[4]{
	\begin{figure}[!ht]
		\ySubFig{#1}\ySubFig{#2}\caption{#3 \label{#4}}
	\end{figure}
}

\newcommand{\yFigThree}[5]{
	\begin{figure}[!ht]
		\ySubFig{#1}\ySubFig{#2}\ySubFig{#3}\caption{#4 \label{#5}}
	\end{figure}
}

\newcommand{\yFigFour}[6]{
	\begin{figure}[!ht]
		\ySubFig{#1}\ySubFig{#2}\ySubFig{#3}\ySubFig{#4}\caption{#5 \label{#6}}
	\end{figure}
}

\newcommand{\ySubFigTone}[1]{
	\begin{subfigure}[t]
		{0.65\textwidth}
		\includegraphics[width=\textwidth]{./#1}
	\end{subfigure}
}

\newcommand{\ySubFigTtwo}[1]{
	\begin{subfigure}[t]
		{0.33\textwidth}
		\includegraphics[width=\textwidth]{./#1}
	\end{subfigure}
}

\newcommand{\yFigFourMine}[6]{
	\begin{figure}[!ht]
		\ySubFig{#1}\ySubFig{#2}\ySubFigTone{#3}\ySubFigTtwo{#4}\caption{#5 \label{#6}}
	\end{figure}
}

\newcommand{\quotes}[1]{``#1''}

\begin{document}
	
\begin{center}
	{\large{\textbf{Gyrokinetic investigation of the damping channels of Alfv\'en modes in ASDEX Upgrade}}}\\
	\vspace{0.2 cm}
	{\normalsize 
		\underline{F. Vannini}$^{1}$, A. Biancalani$^1$, A. Bottino$^1$,\\T. Hayward-Schneider$^{1}$, Ph. Lauber$^1$, A. Mishchenko$^2$, I. Novikau$^{1}$,  E. Poli$^{1}$ and the ASDEX Upgrade team$^1$.
	}\\
$^1$Max-Planck-Institut f\"ur Plasmaphysik, 85748 Garching, Germany \\
$^2$Max-Planck-Institut f\"ur Plasmaphysik, 17491 Greifswald, Germany \\
	\vspace{0.2 cm}
	\small{francesco.vannini@ ipp.mpg.de\\
		}
\end{center}

\begin{abstract}
The linear destabilization and nonlinear saturation of energetic-particle driven Alfv\'enic instabilities in tokamaks strongly depend on the damping channels. In this work, the collisionless damping mechanisms of Alfv\'enic modes are investigated within a gyrokinetic framework, by means of global simulations with the particle-in-cell code ORB5, and compared with the eigenvalue code LIGKA and reduced models. In particular, the continuum damping and the Landau damping (of ions and electrons) are considered. The electron Landau damping is found to be dominant on the ion Landau damping for experimentally relevant cases. As an application, the linear and nonlinear dynamics of toroidicity induced Alfv\'en eigenmodes and energetic-particle driven modes in ASDEX Upgrade is investigated theoretically and compared with experimental measurements.
\end{abstract}

\newpage

\section{Introduction}
In burning plasmas relevant for magnetic fusion energy (MFE) research, an important role is played by energetic particles (EPs). With the term EPs, we refer to fusion reaction products (like alpha particles) or super-thermal ions or electrons resulting from plasma heating. Such particles possess higher velocities compared to those typical of the background plasma. In typical tokamaks the frequency associated with the gyrocenter motion of the EPs falls inside the magnetohydrodynamic (MHD) domain ($O(10^{-2}\,\Omega_{ci})$), being $\Omega_{ci}$ the ion cyclotron frequency. Because of that, through resonant wave-particle interactions, one of the three solution satisfying the MHD dispersion relation can be excited. Among them, the most detrimental are the shear Alfv\'en waves (SAWs), having a group velocity parallel to the equilibrium magnetic field and satisfying the dispersion relation: $\omega=k_{||}\,v_{A}$, where the Alfv\'en speed has been defined $v_{A}=\sqrt{B/(4\,\pi\,\rho_{m0})}$ (being $\rho_{m,0}$ the background plasma mass density and B the background magnetic field strength). The excitation of these modes creates a  transport channel for the EPs, which can lead to loss of EPs before their thermalization, causing a less effective heating and also possibly damaging the vessel of the machine. they are also believed to be responsible of large abrupt events (ALE) observed in the Japanese tokamak (JT-60U) (see Ref.\cite{[Bierwage-2018]}). We can simply outline the whole zoology of SAWs basically in discrete Alfv\'en eigenmodes (AEs) and energetic particle continuum modes (EPMs), (non-normal modes of the SAW continuum spectra, merging as discrete fluctuations at the frequency that maximizes the wave-EP power exchange, above the threshold condition of the continuum damping \cite{[Chen-16]}). 

In this paper the attention will be mainly focused on toroidal Alfv\'en eigenmodes (TAE, Alfv\'en eignemodes lying in the frequency gap caused by the tokamak toroidicity) and EPM. The main goal of this paper will be to understand what are the damping mechanisms of the modes of interest comparing (when possible) the simulations results with the prediction of MHD theory. The proper domain to take into account all the nonlinear effects, like wave-wave and wave-particle interaction, as well as finite-Larmor-radius and finite-orbit-width effects, is represented by the gyrokinetic theory. Because of that, the simulations have been principally carried with the global, nonlinear, electromagnetic, gyrokinetic, PIC code ORB5 \cite{[Jolliet-07],[Bottino-11]} whose model, if properly set, contains the MHD equations as a subset. Some comparisons with the linear gyrokinetic code LIGKA \cite{[LIGKA]} have also been performed.

The paper is structured as follows. In Section 2 a description of the model implemented in the code ORB5 is given. Section 3 and Section 4 will be dedicated to the description of two different damping mechanisms affecting the SAWs: the continuum damping and the Landau damping. They will be briefly explained analytically, starting from the MHD equations. The theory in use will be compared with the results from the simulations carried with ORB5. In Section3 the numerical simulations will be performed in the cylinder limit using simplified profiles. In the simulations in Section 4, a small but finite inverse aspect ratio will be considered, using the equilibrium profiles of the International Tokamak Physics Activity (ITPA, see Ref.\cite{[Koenies_2018]}). In Section 5 the  studies on the linear and nonlinear growth rate and frequency spectra conducted considering experimental profiles from the NLED-AUG case \cite{[Lauber]} will be presented. Finally, summary and future outlook will be shown in Section 6.

\clearpage

\section{Model}

Since the Alfv\'en waves have a frequency much smaller then the typical ion cyclotron frequency ($\Omega_{ci}$) and their amplitude in the core is small compared to the background quantities, a good description of their propagation and interaction with the bulk plasma can be given through the gyrokinetic theory. This allows to retain a kinetic description of the events under consideration, reducing the 6D problem to a 5D one, by averaging the fast gyromotion. In this way the numerical costs are sensibly reduced.

ORB5 is a global, nonliner, gyrokinetic, electromagnetic, PIC code, which can take into account collisions and sources  \cite{[Jolliet-07],[Lanti-2019]}. The gyrokinetic model of ORB5 \cite{[Tronko-2016]} contains the reduced MHD equations as a subset \cite{[Myato-2013]}.
In this section we give a brief description of the gyrokinetic model implemented in ORB5 and briefly show how the implemented equations are solved. We refer for more exhaustive explanations to Ref.\cite{[Lanti-2019]}, which also give a more complete description of the recent updates in ORB5.
Concerning the magnetic equilibrium  in  use, ORB5 can either upload an ideal-MHD equilibrium (solution of the Grad-Shafranov equation) from the CHEASE code \cite{[CHEASE]}, or consider ad-hoc equilibria constituted by circular, concentric magnetic surfaces. It deals with a straight-field line set of coordinates. The magnetic surfaces are labeled by $s=\sqrt{\psi/\psi_{edge}}$, which plays the role of radial coordinate. Here $\psi$ is the poloidal magnetic flux function. The angular dependence is given by the toroidal coordinate $\varphi$ and by the poloidal magnetic angle: 
\begin{equation}
\chi=\frac{1}{q(s)}\int_{0}^{\theta}d\theta^{'} \frac{\vec{B}\cdot\grad{\varphi}}{\vec{B}\cdot\grad{\theta^{'}}}
\end{equation}
being $\theta^{'}$ the geometrical poloidal angle and $q(s)$ the safety factor, defined as:
\begin{equation}
q(s)=\frac{1}{2\pi}\int_{0}^{2\,\pi}d\theta^{'} \frac{\vec{B}\cdot\grad{\varphi}}{\vec{B}\cdot\grad{\theta^{'}}}\quad .
\label{Eq:safety_factor_profile}
\end{equation}
All the quantities in the code are normalized through four reference parameters: the ion mass ($m_{i}$), the ion charge ($q_{i}=e\,Z_{i}$, being $e$ the electric charge and $Z_{i}$ the atomic number), the value of the magnetic field strength on axis ($B_{0}=\abs{\vec{B}(s=0)}$) and the value of the electron temperature at a specified reference position $s_{0}$, $T_{e}(s_{0})$. All other normalized quantities are obtained through these: the time units are provided in the inverse of the ion-cyclotron frequency, $\Omega_{ci}=q_{i}\,B_{0}/(m_{i}\,c)$, the velocity units are normalized through the ions sound velocity ($c_{s}=\sqrt{q_{s}\,T_{e}(s_{0})/m_{i}}$, being the temperature measured in $keV$), the length units through the ion sound Larmor radius ($\rho_{s}=c_{s}/\Omega_{ci}$) and the densities are normalized by means of their average in space.
The Vlasov-Maxwell gyrokinetic equations are derived through variational principles from a discrete gyrokinetic Lagrangian. This choice allows to take into account all the simplification needed in the model under consideration, directly in the Lagrangian and then derive the gyrokinetic equation. An immediate consequence is that it is possible to consistently derive conserved quantities (like the energy, see  Ref.\cite{[Bottino-2015]}), that are also used in ORB5 to test the quality of the simulations performed. This choice also allows to derive naturally the weak gyrokinetic form of the field equations. In the Lagrangian an ordering is present, separating the effects given from the geometry of the non-uniform magnetic field, from those related to the fluctuations of the electromagnetic perturbation. This means that (as can be derived, see Ref.\cite{[Tronko-2016],[Tronko-2017]}) the small parameter related to the variation of the background magnetic field $\epsilon_{B}=\rho_{th}L_{B}$, (being $\rho_{th}$ the thermal Larmor radius and $L_{B}$ the typical variation length of the magnetic field) and the small parameter related to the fluctuating electromagnetic field ($\epsilon_{\delta}$) are related through:
		\begin{equation}
		\epsilon_{B}=O(\epsilon_{\delta}^{2})\quad .
		\end{equation}
In this way the action functional, written in \quotes{$p_{z}$-formulation} appears to be the following:
\begin{equation}
\begin{multlined}
A=\int_{t_{0}}^{t_{1}} L\,dt =\sum_{s}^{}\int dt\,d\Omega \left(\frac{q_{s}}{c}\vec{A^{*}}\cdot\vec{\dot{X}}+\frac{m_{s}\,c}{q_{s}}\mu\,\dot{\theta}-H_{0}\right)f_{s} +\\
-\epsilon_{\delta}\sum_{s\not= e}^{}\int dt\,d\Omega\,H_{1}\,f_{s}-\epsilon_{\delta}\int dt\,d\Omega\,H_{1}^{dk}\,f_{e}+\\
-\epsilon_{\delta}^{2}\sum_{s\not= e}^{}\int dt\,d\Omega\,H_{2}\,f_{eq,s}-\alpha\epsilon_{\delta}^{2}\int dt\,d\Omega\,H_{2}^{dk}\,f_{eq,e}-\alpha\epsilon_{\delta}^{2}\int dt\,dV\,\frac{\abs{\grad_{\perp}{A_{1,||}}}^{2}}{8\pi}
\end{multlined}
\label{Eq:Action_ORB5}
\end{equation}
where $\alpha=0$ gives the electrostatic model, while $\alpha=1$ the electromagnetic one. In Eq.\ref{Eq:Action_ORB5} $d\Omega=dV\,dW$, being $dW=B^{*}_{||}d\mu\,dp_{z}$. A sum over the species \quotes{$s$} also appears. The symplectic magnetic field is defined through the symplectic magnetic potential $\vec{A}^{*}=\vec{A}+(c/q_{s})p_{z}\uvec{b}$ being $\uvec{b}$ the unitary vector parallel to the background magnetic field. The canonical gyrocenter momentum is $p_{z}=m_{s}v_{\parallel}+\alpha\epsilon_{\delta}(q_{s}/c)\,A_{1\parallel}$. In the action functional, some approximations have been done. The quasi-neutrality allows to consider in Eq.\ref{Eq:Action_ORB5} only the contribution given from the magnetic potential, neglecting the one given from the perturbed electric field. Also the incompressibility of the parallel perturbed magnetic field is assumed $B_{1,||}=o(B_{1,\perp})$ and only the perpendicular component of the perturbed magnetic potential is retained:$B_{1,||}=\nabla\cross (A_{1,||}\vec{b})\sim \nabla A_{1,||}\cross \vec{b}$.
In Eq.\ref{Eq:Action_ORB5} it must be noted that while $H_{0}$,$H_{1}$ multiply the total distribution functions $f_{s},f_{e}$, $H_{2}$ is related only to the equilibrium distribution function. Thanks to this choice nonlinear second order terms do not appear in the gyrocenter dynamics and the field equations are linear. The gyrocenter hamiltonians appearing are:
\begin{align}
H_{0}=\frac{p_{z}^{2}}{2\,m_{s}}+\mu\,B && H_{1}=q_{s}\left\langle\phi_{1}-\alpha A_{1,\parallel}\frac{p_{z}}{m_{s}\,c}\right\rangle \end{align}
\begin{align*}
H_{2}=-\frac{m_{s}c^{2}}{2\,B^{2}}\abs{\grad_{\perp}{\phi_{1}}}^{2}+\alpha\frac{q_{s}^{2}}{2m_{s}\,c^{2}}\left\langle A_{1,\parallel}\right\rangle^{2}
 \end{align*}
 where the gyroaveraging operator has been introduced $\left\langle f\right\rangle =\frac{1}{2\pi}\int_{0}^{2\pi}d\theta f$. The gyroaveraging is removed for the electrons that are treated as drift-kinetic:
 \begin{align}
     H_{1}^{dk}=-e\left(\phi_{1}-\alpha A_{1,\parallel}\frac{p_{z}}{m_{s}\,c}\right) && H_{2}^{dk}=\alpha\frac{e^{2}}{2m_{e}c^{2}}A_{1,\parallel}^{2}
 \end{align}
For the distribution of the species $s$ the linear gyrokinetic Vlasov equation is:
\begin{equation}
\derivative{f_{s}}{t}=\frac{\partial f_{s}}{\partial t}+\vec{\dot{X}}\cdot\grad{f_{s}}+\dot{p}_{z}\frac{\partial f_{s}}{\partial p_{z}}=0
\end{equation}
where the gyrokinetic characteristics can be derived from Eq.\ref{Eq:Action_ORB5} and are:
\begin{equation}
\begin{cases}
\begin{multlined}
\vec{\dot{X}}=\frac{c\uvec{b}}{q_{s}B^{\star}_{\parallel}}\cross\grad{H}+\frac{\partial H}{\partial p_{z}}\frac{\vec{B^{\star}}}{B^{\star}_{\parallel}}
\end{multlined}
\\
\\
\begin{multlined}
\dot{p}_{z}=-\frac{\vec{B^{\star}}}{B^{\star}_{\parallel}}\cdot\grad{H}
\end{multlined}
\end{cases}
\end{equation}
The field equations, quasineutrality and Ampère, are both derived from Eq.\ref{Eq:Action_ORB5} via functional derivatives on the perturbed field. ORB5 splits the total distribution function in a background distribution function $f_{0}$ and in a time dependent one $\delta f$ and discretize this latter through numerical particles (markers) used to sample the phase space. Through an operator splitting approach the code solves first the conlisionless dynamics (using a 4th-order Runge-Kutta method) and then treats the collisions with a Lanngevin approach.  The quasineutrality and Ampère equations are solved using the Galerkin methods and the perturbed fields are discretized through cubic B-splines finite elements defined on a grid $(N_{s},N_{\chi},N_{\phi})$. Finally it is important to mention that from the numerical side, recently the mixed-representation (\quotes{pullback} scheme \cite{[Mishchenko-PB]}) has solved the so-called \quotes{cancellation problem} for electromagnetic simulations.

\clearpage

\section{Continuum damping }

In the present section,  the continuum damping will be studied.  The tokamak configuration will be selected in order to have the continuum damping as main damping mechanism affecting the Alfv\'en waves. In order to do so, it is first important to understand what are the equations governing the Alfv\'en waves. These will be obtained under the validity of the ideal magnetohydrodynamic (MHD) theory by treating MHD equations with a perturbative approach. 
The Alfv\'en wave's dynamics can be expressed starting from the quasi-neutrality condition $\grad\cdot{\vec{\delta J}}=0$, (being $\vec{\delta J}$ the perturbed current) that rewritten in terms of its components parallel and perpendicular to the background magnetic field ($\uvec{b}=\vec{B}/B$) reads:
\begin{equation}
\grad\cdot\vec{\delta J_{\perp}}+\vec{B}\cdot\grad{\frac{\delta J_{\parallel}}{B}}=0 \quad .
\label{Eq:quasi_neutrality_MHD}
\end{equation}
Following Ref.\cite{[Zonca_Vlad_99]}, in order to obtain a simplified but  relevant set of equations, modes with $k_{\perp}\gg k_{\parallel}$ are considered, so that the time scale between incompressible shear Alfv\'en waves and compressional waves can be separated. To further simplify the problem, we consider a pressureless plasma ($P=0$)  obtaining the following vorticity equation:
\begin{equation}
\vec{B}\cdot\grad\left[\frac{1}{B}\grad^{2}_{\perp}\left(\frac{1}{B}\vec{B}\cdot\grad\,\delta\phi\right)\right]-\grad\cdot\left(\frac{1}{v_{A}^{2}}\frac{\partial^{2}}{\partial t^{2}}\grad_{\perp}\delta\phi\right)=0 \quad .
\label{Eq:vorticity_pressureless}
\end{equation}
A differential equation for the perturbed scalar potential $\delta\phi$ is thus obtained. It is linked to the perturbed magnetic potential ($\vec{\delta A}\approx\delta A\,\uvec{b}$) through the condition $\delta E_{\parallel}=0$, derived from the ideal Ohm's law. In this section a non-uniform plasma equilibrium with cylindrical limit, will be considered. $a$ will denote the typical length scale perpendicular to the equilibrium magnetic field while $R_{0}$ will represent the typical length scale parallel to it. The equilibrium magnetic field, in a coordinate system $(r,\theta,z)$ will be assumed to be $\vec{B}=(0,B_{0,\theta}(r),B_{0,z}(r))$.
The geometrical radius $r$ can be used in the cylindrical limit instead of the flux coordinate $s$. 
By assuming a shear Alfv\'en oscillation of the scalar potential $\delta\phi(r,\theta,\phi,t)$ of the form:
\begin{equation}
\delta\phi=\sum_{m,n}\delta\phi_{m,n}(r)\,e^{i(m\theta-\frac{n\,z}{R_{0}}-\omega t)}\quad ,
\end{equation}
where $m$ is the poloidal mode number, we can now write Eq.\ref{Eq:vorticity_pressureless} in cylindrical coordinates:
\begin{equation}
\frac{1}{r}\frac{\partial}{\partial r}r\left[\left(\frac{m}{q(r)}-n\right)^{2}+\frac{R_{0}^{2}}{v_{A}^{2}}\frac{\partial^{2}}{\partial t^{2}}\right]\frac{\partial}{\partial r}\left(\frac{\delta\phi}{r}\right)=\frac{m^{2}}{r^{2}}\left[\left(\frac{m}{q(r)}-n\right)^{2}+\frac{R_{0}^{2}}{v_{A}^{2}}\frac{\partial^{2}}{\partial t^{2}}\right]\delta\phi \quad ,
\label{Eq:vorticity_pressureless_cylindrical}
\end{equation}
where the local safety-factor profile has been defined:
\begin{equation}
q(r)=\frac{r\,B_{0,z}}{R_{0}\,B_{0,\theta}} \quad .
\end{equation}
The shear Alfv\'en wave dispersion relation is then found to be:
\begin{equation}
\omega^{2}_{A}=v_{A}^{2}\,k_{m,n}^{2}=\frac{v_{A}^{2}}{R_{0}^{2}}\left(\frac{m}{q(r)}-n\right)^{2}
\label{Eq:Dispersion_relation_cylindrical}
\end{equation}
Equation \ref{Eq:Dispersion_relation_cylindrical} proves that the shear Alfv\'en waves are local plasma oscillations, having a frequency spectrum that varies continuously throughout the plasma radial direction. Due to the hypothesis $k_{\perp}\gg k_{\parallel}$ the local nature of the continuum plasma oscillation can be exploited by reducing Eq.\ref{Eq:vorticity_pressureless_cylindrical} to:
\begin{equation}
\begin{multlined}
\frac{1}{r}\frac{\partial}{\partial r}r\left[\left(\frac{m}{q(r)}-n\right)^{2}-\omega^{2}\frac{R_{0}^{2}}{v_{A}^{2}}\right]\frac{\partial\delta\phi}{\partial r}=0 \quad ,
\end{multlined}
\end{equation}
which integrated in the radial domain, becomes a differential equation for the radial electric field $E_{r}$:
\begin{equation}
\begin{multlined}
\left(\omega_{A}^{2}+\frac{\partial^{2}}{\partial t^{2}}\right)E_{r}=0\quad \Rightarrow\quad 
E_{r}=E_{0}\,e^{-i\,\omega_{A}(r)\,t} \quad .
\end{multlined}
\end{equation}
Assuming now a dispersion relation of the form $\omega_{A}(r)=\omega_{A0}+\omega^{\prime}_{A}\,(r-r_{0})$ and by Fourier transforming the radial electric field in the radial coordinate the following relation is obtained:
\begin{equation}
(\mathscr{F}E_{r})(k_{r})=\sqrt{2\pi}\,E_{0}\,e^{-i\left(\omega_{A0}-\omega^{\prime}r_{0}\right)t}\delta(k_{r}+\omega_{A}^{\prime}\,t)\quad\quad k_{r}\propto -\omega_{A}^{\prime}\,t
\label{Eq:continuum_damping_wave_number}
\end{equation}
The obtained linear dependence in time of the radial wave number is a proof of the phase mixing together with the fact that, being $E_{r}(r,t)=-i\,k_{r}(t)\phi(r,t)$, the scalar potential exhibits the characteristic decay called continuum damping:
\begin{equation}
\delta\phi\propto \abs{\omega_{A}^{\prime}\,t}^{-1}
\label{Eq:continuum_damping_scalar_potential_decay}
\end{equation}
as it was proved in Ref.\cite{[Zonca_Vlad_99]} (see also Ref.\cite{[Zonca_2008],[Palermo_2016],[Biancalani_2016]} for the application to Geodesic Acoustic modes, GAMs). To see evidence of this mechanism, the simulations presented in this section have been run with simplified geometries and profiles, without the presence of EPs. In order to be in the cylinder limit the inverse aspect ratio has been chosen to be $\epsilon=0.01$ and flat density ($n_{e}=n_{i}=2.22\,\cdot10^{20}\,m^{-3}$) and temperature ($T_{e}=T_{i}=0.01\,keV$) profiles have been taken into account. This leaves all the radial dependence of the dispersion relation in the safety factor profile. Moreover, in this temperature regime the Landau damping can be neglected \cite{[Chen_1974]}. In the simulations under examination only one axysimmetric perturbation $n=0\,,\,m=1$ peaked at the radial position $r = 0.6$, has been considered, together with a linear safety-factor profile $q=q_{0}+q_{1}\cdot r$ so that: $\omega_{A}=\frac{v_{A}}{R_{0}}\frac{1}{q_{0}+q_{1}\,r}$. The other important parameters in the simulations (minor and major radius, value on axis of the equilibrium magnetic field, ion cyclotron frequency, Alfv\'en frequency on-axis and the ratio of the last two), are reported in Tab.\ref{tab:continuum_damping_details}.

\begin{table}[htbp]
	\centering
	\caption{Main simulation's parameters of the reference case for the study of the continuum damping.}
	\vspace{1ex}	
	\label{tab:continuum_damping_details}
	\begin{tabular}{||c|c|c|c|c|c|c|c|c||}\hline
		
		\textbf{$a_{0}\,[m]$} & \textbf{$R_{0}\,[m]$}&\textbf{$B_{0}\,[T]$} &\textbf{$\Omega_{ci}\,[rad/s]$}&\textbf{$\omega_{A0}\,[rad/s]$}&\textbf{$\Omega_{ci}/\omega_{A0}$}\\
		\hline

		$0.1$ & $10$&$3$&$2.87\cdot 10^{8}$&$4.38\cdot 10^{5}$&$655$ \\

		\hline
		
		
	\end{tabular}
	\label{tab:1}
\end{table}

In Fig.\ref{fig:continuum_damping_q1_05} on the left, the measured values for the wave numbers $k_{r}$ are shown for a simulation having $q_{0}=1.75$ and $q_{1}=0.5$. They have been measured interpolating the mode structure with a sinusoidal function at times where a maximum has been reached at the radial position $r = 0.6$. By linearly inetrpolating the measured wave numbers it is possible to calculate the coefficient $k_{r,1}$, which is found to be in reasonable agreement with the theoretical expectations Eq.\ref{Eq:continuum_damping_wave_number}. In Fig.\ref{fig:continuum_damping_q1_05} the dynamic of the scalar potential at some radial positions is shown, together with the predicted decay, Eq.\ref{Eq:continuum_damping_scalar_potential_decay}.

\yFigTwo
{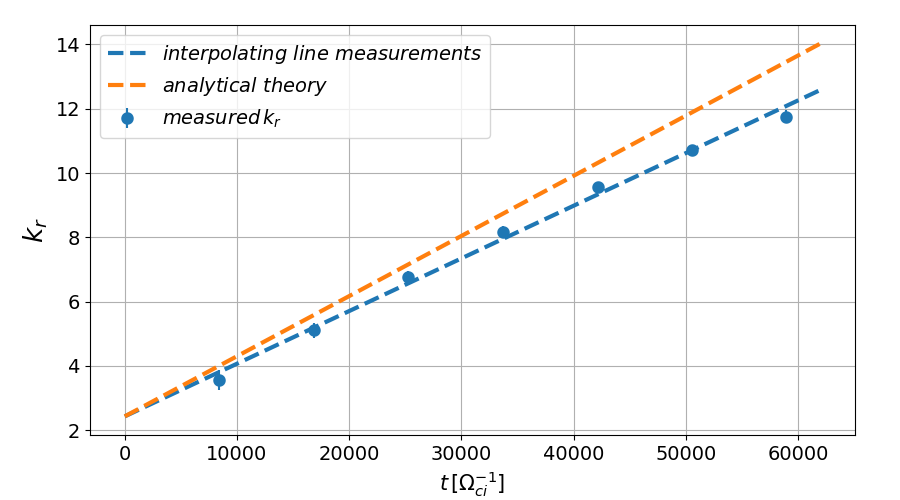}
{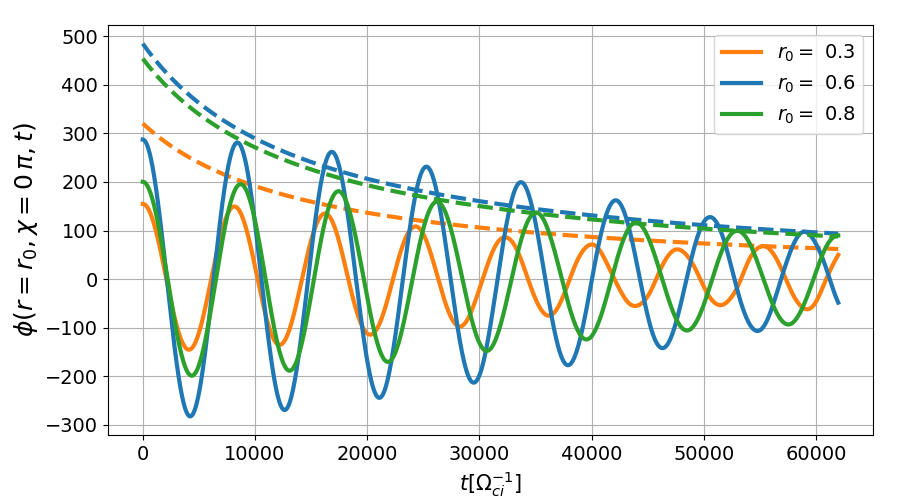}
{Left: Radial wave number dependence on time. The results of ORB5 are given by dots. The theoretical prediction for this simulation ($q_{0}=1.75$ and $q_{1}=0.5$) is that $k_{r,1}=0.123\,\omega_{A0}$, while the measured value is $k_{r,1}=0.107\,\omega_{A0}$. Right: Perturbation amplitude dependence on time. Analytical estimation are given by the dashed lines (curves decaying in time as $\phi\sim\abs{\omega_{A}^{\prime}\,t}^{-1}$).  The scalar potential measured at different radial positions is given by continuous lines. No EPs are present here.} 
{fig:continuum_damping_q1_05}

\begin{figure}[h!]
	\begin{center}
		\vskip -0.4em
		\includegraphics[width=0.6\textwidth]{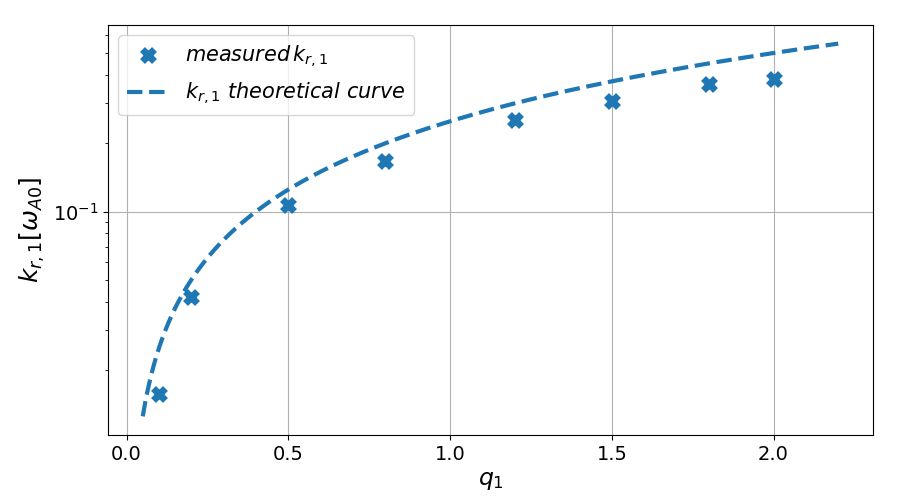}
		\vskip -1em
		\caption{Dependence of $k_{r,1}$ on the slope of the safety factor profile. Analytical estimation are given by the dashed lines and the results of ORB5 are given by dots. No EP are present here.} 
		\label{fig:continuum_damping_q1}
	\end{center}
\end{figure}

In Fig.\ref{fig:continuum_damping_q1} finally the obtained values of the coefficients $k_{r,1}$ have been plotted against different values of the slope of the safety factor profiles ($q_{1}$) in use in the different simulations and compared with Eq.\ref{Eq:continuum_damping_wave_number}. Given the reasonable agreement found between the results of the numerical simulations and the theory, we can say to have verified the relevance of the continuum damping as main damping mechanism for this specific case and to have observed the presence of phase mixing.

\clearpage
\section{Landau damping}
In the present section bulk species temperatures high enough to make the continuum damping negligible with respect to the Landau damping will be considered, so that the latter becomes the main damping mechanisms.

The  attention will be focused also on a particular Alfv\'en eigenmode, the toroidal Alfv\'en eignemode (TAE). Its characteristic frequency lies in the gap created in the continuum spectra by two close poloidal modes $(m,m+1)$ which are coupled because of the finite tokamak toroidicity, \cite{[Fu_1989]}. A TAE is located at a radial  position  $r_{0}$ satisfying: $
q(r_{0})=\frac{2\,m+1}{2\,n}$. 
The theoretical derivation exposed in Ref.\cite{[Fu_1989],[Koles_2002]} will be now followed. Here,  a kinetic transverse part of the wave-induced current $\vec{\delta J_{\perp}^{k}}$ is added to the ideal MHD current, so that Eq.\ref{Eq:quasi_neutrality_MHD} becomes:
\begin{equation}
\grad\cdot(\vec{\delta J^{MHD}}+\vec{\delta J_{\perp}^{k}})=0\quad .
\label{Eq:quasi_neutrality_correct}
\end{equation}
Equation \ref{Eq:quasi_neutrality_correct} is then multiplied by $\delta\phi$ and integrated in the overall plasma volume obtaining:
\begin{equation}
\int d\vec{x}\,\vec{\delta J^{MHD}}\cdot\grad{\delta\phi}+\int d\vec{x}\,\vec{\delta J_{\perp}^{k}}\cdot\grad_{\perp}{\delta\phi}=0
\label{Eq:quasi_neutrality_correct_1step}
\end{equation}
where it was assumed as boundary conditions $\int d\vec{x}\cdot\vec{\delta J}\delta\phi=0$. Calling $\omega_{0}$ the frequency of the wave solution of the ideal MHD vorticity equation, we can consider $\omega=\omega_{0}+\delta\omega$ the solution of the new vorticity equation Eq.\ref{Eq:quasi_neutrality_correct}, being $\delta\omega\ll\omega_{0}$ . Following a perturbative approach, an expression for $\gamma=\Im{\omega}$ is obtained from Eq.\ref{Eq:quasi_neutrality_correct_1step}:

\begin{equation}
\begin{multlined}
\gamma=\frac{2\pi }{c^{2}}\frac{\sum_{m,n}\int_{V}^{}d^{3}x\,\vec{\delta J_{\perp m,n}^{k}}\cdot\grad\delta\phi^{*}_{m,n}}{\sum_{m,n}\int_{V}^{}d^{3}x\frac{1}{v_{A}^{2}}\left[\abs{\delta\phi'_{m,n}}^{2}+\left(\frac{m}{r}\right)^{2}\abs{\delta\phi_{m,n}}^{2}\right]}
\end{multlined}
\label{Eq:Growth_rate_general_formula}
\end{equation}
where all the appearing perturbed quantities have been decomposed in  Fourier components in the poloidal plane.
In order to obtain a simplified equation for $\gamma$, some further calculations have been done and will be now described. Eq.\ref{Eq:Growth_rate_general_formula} is then written in cylindrical coordinates, after writing the perturbed current in terms of the perturbed distribution function and this in terms of the unperturbed distribution function $F_{0}$. Assuming a Maxwellian distribution function $F_{0}$ and focusing our attention on TAE (that is assuming to have a perturbation $\delta\phi$ strongly peaked at the radial position where we expect to have a TAE), we obtain:

\begin{equation}
\gamma=\sum_{j}\gamma_{j}\quad\quad \gamma_{j}
= -\beta_{j}\, \,q_{0}^{2}\,\frac{v_{A}}{2\,q_{0}\,R_{0}}\,\left[G_{m\,j}+n\,q_{0}\,r_{L\theta,j}\frac{1}{n_{0,j}}\frac{\partial n_{0,j}}{\partial r}\left(H_{m\,j}+\eta\,J_{m\,j}\right)\right]
\quad .
\label{Eq:Eq_final_Betti}
\end{equation}
Being:
\begin{align}
\Omega_{\theta,j}=\frac{e\,B_{p}}{m_{j}\,c} \quad r_{L\theta}=\frac{v_{th,j}}{\Omega_{\theta,j}} \quad \beta_{j} = 8\pi\frac{n_{0,j}\,T_{j}}{B_{0}^{2}} \quad \eta_{j}=\frac{\partial\log(T_{j})}{\partial\log(n_{0,j})}\quad \lambda_{j}=v_{A}/v_{th,j}\quad .
\end{align}
And:
\begin{equation}
\begin{cases}
\begin{multlined}
g_{m,j}(\lambda_{j})= \frac{\pi}{2}\lambda_{j}(1+2\lambda_{j}^{2}+2\lambda_{j}^{4})e^{-\lambda_{j}^{2}}\quad\quad G_{m\,j}=g_{m,j}(\lambda_{j})+g_{m,j}(\lambda_{j}/3)
\end{multlined}
\\
\\
\begin{multlined}
h_{m,j}(\lambda_{j})= \frac{\pi}{2}(1+2\lambda_{j}^{2}+2\lambda_{j}^{4})e^{-\lambda_{j}^{2}}\quad\quad  H_{m\,j}=h_{m,j}(\lambda_{j})+\frac{1}{3}h_{m,j}(\lambda_{j}/3)
\end{multlined}
\\
\\
\begin{multlined}
j_{m,j}(\lambda_{j})= \frac{\pi}{2}\left(\frac{3}{2}+2\lambda_{j}^{2}+\lambda_{j}^{4}+2\lambda_{j}^{6}\right)e^{-\lambda_{j}^{2}}\quad\quad J_{m\,j}=j_{m,j}(\lambda_{j})+\frac{1}{3}j_{m,j}(\lambda_{j}/3)\quad .
\end{multlined}
\end{cases}
\end{equation}
In Eq.\ref{Eq:Eq_final_Betti}, $\gamma$ has been decomposed in the species contributions (the sum over $j$). It is formally identical to the one derived in Ref.\cite{[Betti_1992]}. The difference lies in the fact that in Ref.\cite{[Betti_1992]} the authors have obtained the estimation for $\gamma$ starting from energy principles, while here everything has been done by adding a correction to the MHD quasi-neutrality equation and thus to the Alfv\'en dynamics \cite{[Koles_2016]}. Since we are interested in the study of the Landau damping, we will not consider the EPs contribution, which actually drives the mode unstable. Eq.\ref{Eq:Eq_final_Betti} depends on the ratio between the Alfv\'en speed and the thermal velocity of the considered species. This means that $\lambda_{j}\sim m_{j}^{1/2} $ and, since the bulk ion mass is bigger than the electron mass (for Hydrogens $m_{H}\sim 2000\,m_{e}$) one can understand that the ion contribution is negligible with respect to the electron contributions (because of the presence of the exponential terms in the polinomia). Because of that we will focus our attention on the electron Landau damping in this analytical derivation.

In this section an equilibrium with small, but finite value of inverse aspect ratio will be considered, $\epsilon=0.1$. The temperature profiles are flat. When considered, the fast particles have a density profile peaked on axis (see Fig.\ref{Fig:profiles_ITPA} on the left). 
The magnetic equilibrium and profiles are those of the ITPA-TAE international benchmark case \cite{[Koenies_2018]} and the safety factor profile is shown in Fig.\ref{Fig:profiles_ITPA} on the right. The main results that will be displayed in this section, have been obtained considering heavier electrons: $m_{e}=m_{H}/200$. This has been checked to be at convergence. In Tab.\ref{tab:landau_damping_details} other important details of the simulations are presented.
\begin{table}[htbp]
	\centering
	\caption{Main simulation's parameters of the  ITPA-TAE case.}
	\vspace{1ex}	
	\label{tab:landau_damping_details}
	\begin{tabular}{||c|c|c|c|c|c|c|c|c||}\hline
		
		$a_{0}\,[m]$ & $R_{0}\,[m]$&$B_{0}\,[T]$ &$\Omega_{ci}\,[rad/s]$&$\omega_{A0}\,[rad/s]$&$\Omega_{ci}/\omega_{A0}$\\
		\hline

		$1$ & $10$&$3$&$2.87\cdot 10^{8}$&$1.46\cdot 10^{6}$&$196$ \\

		\hline
		
		
	\end{tabular}
	\label{tab:1}
\end{table}


\yFigTwo
{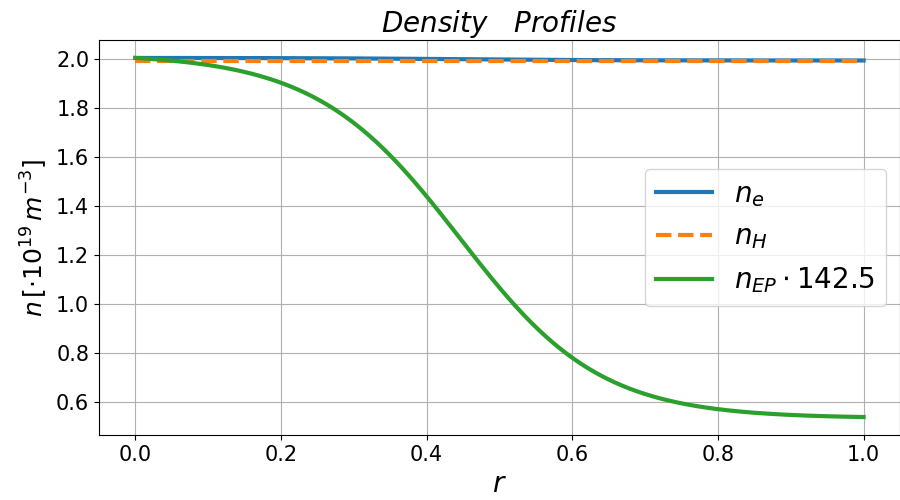}
{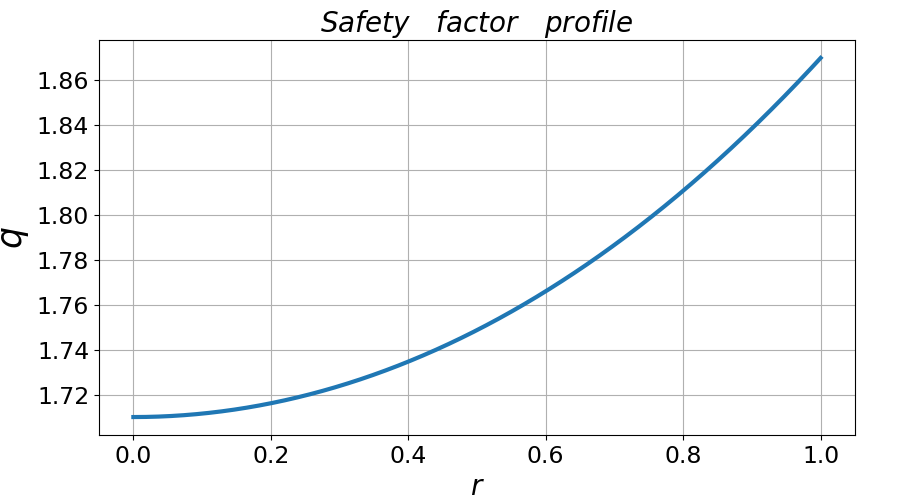}
{Density profiles and safety factor of the ITPA-TAE case ($q(r)\simeq 1.71+0.15 r^{2}$).}
{Fig:profiles_ITPA}

\yFigTwo
{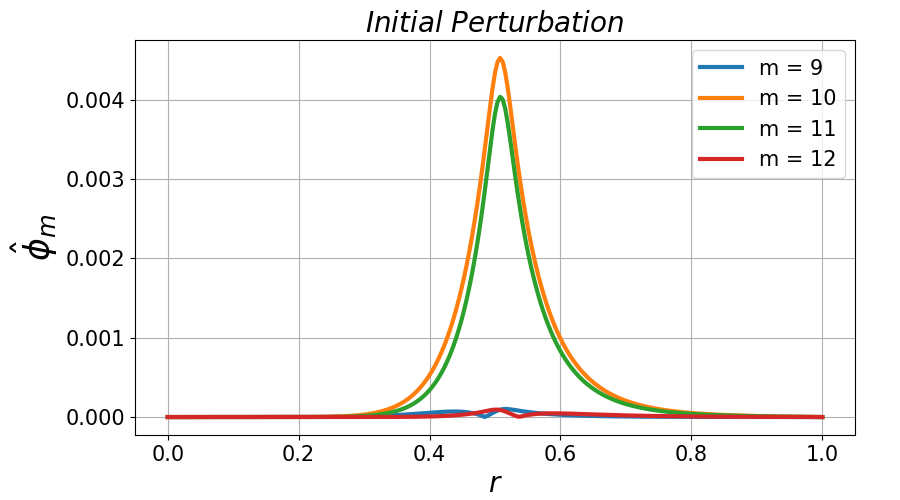}
{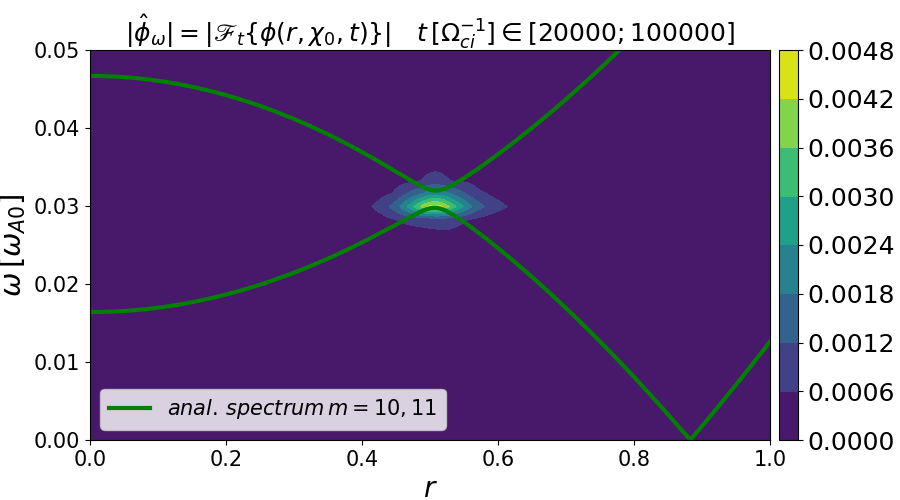}
{Left:Initial mode structure.Right: Frequency spectra without energetic particles.}
{Fig:ITPA_initialization}

\begin{figure}[h!]
	\begin{center}
		\vskip -0.2em
		\includegraphics[width=0.61\textwidth]{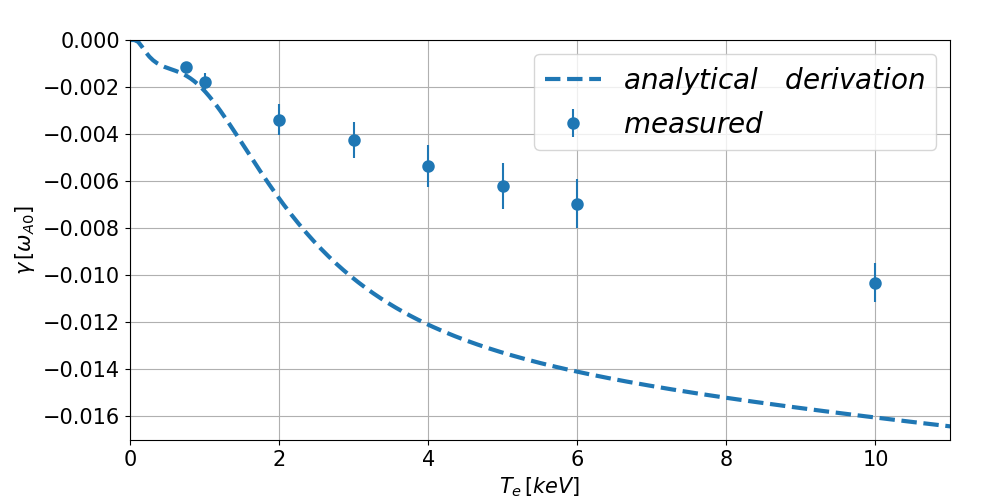}
		\vskip -1em
		\caption{Landau damping dependence versus the electron temperature.} 
		\label{Fig:Electron_landau_damping}
	\end{center}
\end{figure}

The chosen initial potential perturbation is peaked around $r=0.5$ and is constituted by one single toroidal mode number $n=6$ while the poloidal mode numbers $9\leq m\leq 12$ are considered (see Fig.\ref{Fig:ITPA_initialization} on the left). A TAE is located at $r=0.5$ in the gap of the continuum spectra created by the coupling of the poloidal modes $m=10$ and $m=11$, as it is shown in Fig.\ref{Fig:ITPA_initialization} on the right. In Fig.\ref{Fig:Electron_landau_damping} the dependence of the damping rate against the value of the (flat) electron temperature is shown. The damping rate is found to increase with the increasing electron temperature. This is an evidence that the dominant damping is the electron Landau damping. The errorbar of the measured points correspond to $20\%$ of their value. This because, as it is shown in Fig.\ref{Fig:Electron_landau_damping_dependencies} on the left, the damping rate value has a dependence on the chosen width of the perturbation. For completeness in Fig.\ref{Fig:Electron_landau_damping} the approximated analytical electron Landau damping formula is also shown (dashed line). A reasonable qualitative agreement is found between the predicted decay and the simulation results.
Finally in Fig.\ref{Fig:Electron_landau_damping_dependencies} on the right, the dependence of the measured damping rate of ORB5 simulations against the electron mass has been shown.  For decreasing electron masses, the absolute value of the damping rate is shown to decrease, consistently with theory of the electron Landau damping. In summary, it has been proved that the bulk electrons provide the main damping mechanism of the observed Alfv\'en modes in this particular regime. Several approximations have been done in the analytical theory, inter alia only passing particles are considered thus neglecting the contributions of barely trapped electrons, which are thought to be important, and which are included in our numerical simulations.  

\yFigTwo
{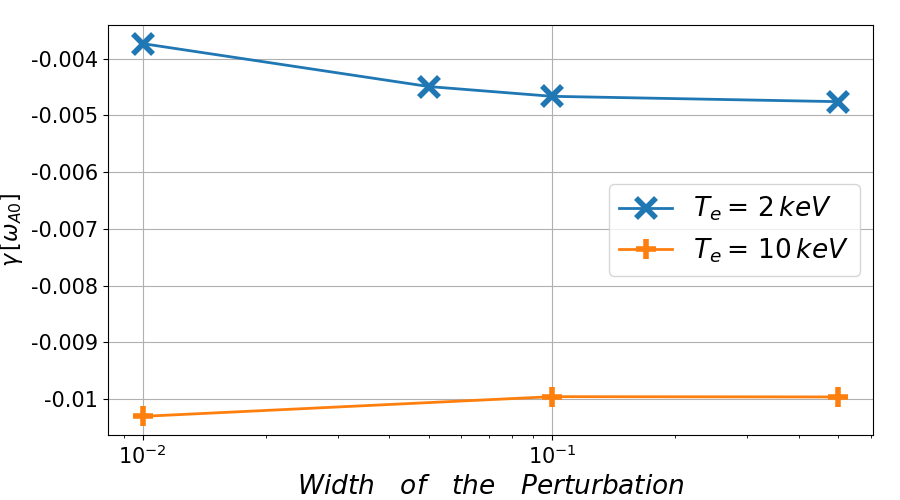}
{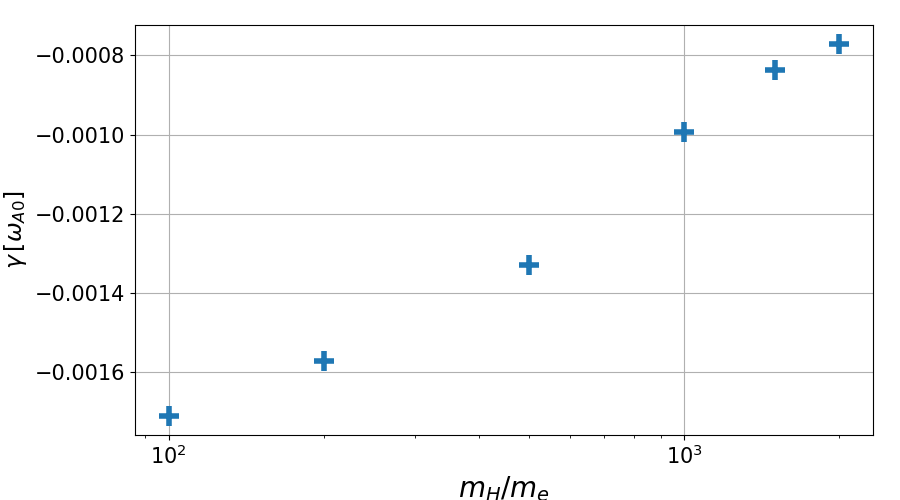}
{Left:Damping rate dependence on the width of the initial Gaussian beam. Right: Damping rate dependence on the electron mass for ORB5 simulations.}
{Fig:Electron_landau_damping_dependencies}

\clearpage
\section{NLED-AUG case}

In the present section the results of numerical simulations involving a realistic scenario will be presented. 

The shot number $\#31213$ of ASDEX-Upgrade (AUG) has been selected within the Non-Linear Energetic-particle Dynamics (NLED) Eurofusion enabling research project \cite{[Lauber]}. Here an early off-axis NBI (with $T_{EP}\sim 93\,keV$) occurs with an injection angle (angle between the horizontal axis and the beam-line) of $7.13^\circ$. The magnetic equilibrium measured at the time $t=0.84\,s$ is considered in the present simulations (see Fig.\ref{Fig:NLED_AUG_magnetic_equilibrium} on the bottom left). This case is referred  to as \quotes{NLED-AUG case}. Further description of this case can be found in Ref.\cite{[Lauber]}. The NLED-AUG case is found to be of great interest  because of its rich linear and nonlinear dynamics arising from the interaction of the modes with the EPs.

Tab.\ref{tab:nled_aug_details_1} contains the details of the main parameters considered in the simulations. Tab.\ref{tab:nled_aug_details_2} shows the values of the bulk species profiles on axis in the absence of EPs.  The bulk ions, as well as the EPs (when considered), are constituted by deuterium. The EP temperature will be always considered to be radially flat and equal to $T_{EP}=113\,keV$. For the EPs density profiles, an off-axis density profile fitting the experimental profiles is considered with Maxwellian distribution function. For comparison we also run simulations with an on-axis EPs density profiles. Note that when EPs are included the electron density profiles is changed in order to match quasi-neutrality $n_{e}=Z_{i}\cdot n_{i}+Z_{EP}\cdot n_{EP}$.

In Fig.\ref{Fig:NLED_AUG_magnetic_equilibrium} the safety factor profile is shown, together with the temperature profiles of the bulk species. The safety factor profile has a reversed shear, with $q_{min}(r = 0.5) \simeq 2.28$. The density profiles in use will be shown in the following subsection. For numerical reasons, the electron mass is chosen to be $m_{e}=m_{D}/500$, being $m_{D}$ the deuterium mass.

This section is divided in two subsections. In the first, the results of numerical simulations only involving the linear dynamics will be discussed. In particular, the dependence of $\gamma$ against the electron temperature will be shown, with and without EPs contribution. Also the results of a benchmark with LIGKA are presented. In the second subsection instead, the results of simulations involving also the nonlinear dynamics  will be presented. Finally, it is important to remind that, unless specifically written, the bulk and energetic ions will be treated as drift-kinetic. The initial perturbation considered will take into account just one toroidal mode number ($n=1$) and the poloidal mode number $0 \leq m \leq 7$.

\begin{table}[htbp]
	\centering
	\caption{Main simulation's parameters of the NLED-AUG case.}
	\vspace{1ex}	
	\begin{tabular}{||c|c|c|c|c|c|c|c|c||}\hline
		
		\textbf{$a_{0}\,[m]$} & \textbf{$R_{0}\,[m]$}&\textbf{$B_{0}\,[T]$} &\textbf{$\Omega_{ci}\,[rad/s]$}&\textbf{$\omega_{A0}\,[rad/s]$}&\textbf{$\Omega_{ci}/\omega_{A0}$}\\
		\hline

		$0.482$ & $1.666$&$2.202$&$1.0539\cdot 10^{8}$&$4.98\cdot 10^{6}$&$21.15$ \\

		\hline
		
		
	\end{tabular}
	\label{tab:nled_aug_details_1}
\end{table}

\begin{table}[htbp]
	\centering
	\caption{Profile's parameters of the NLED-AUG case.}
	\vspace{1ex}	
	\begin{tabular}{||c|c|c|c|c|c||}\hline
		
		$T_{e}(s=0)\,[keV]$&$T_{i}(s=0)\,[keV]$ &$n_{e}(s=0)\,[m^{-3}]$&$n_{i}(s=0)\,[m^{-3}]$&$n_{f}(s=0)\,[m^{-3}]$ \\
		\hline

		$0.709$ &$2.48$  &$1.672\cdot 10^{19}$&$1.6018\cdot 10^{19}$ &$6.98\cdot 10^{17}$\\

		\hline
		
		
	\end{tabular}
	\label{tab:nled_aug_details_2}
\end{table}

\yFigThree
{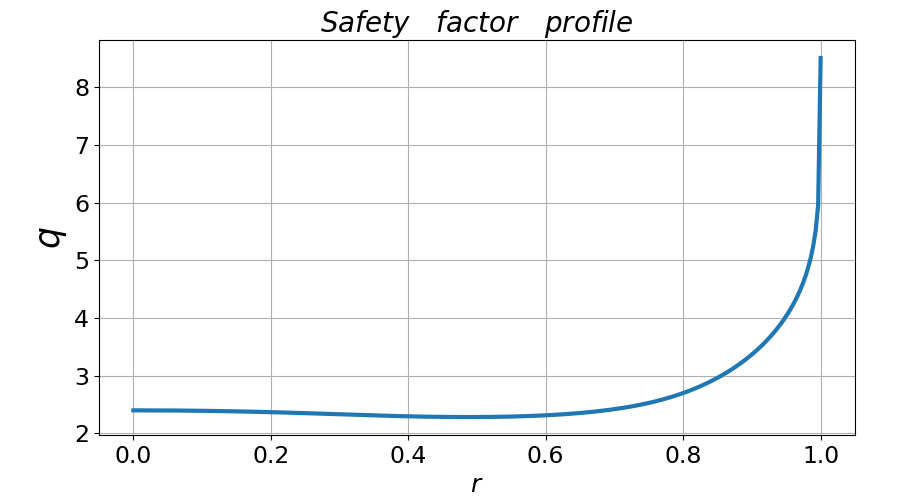}
{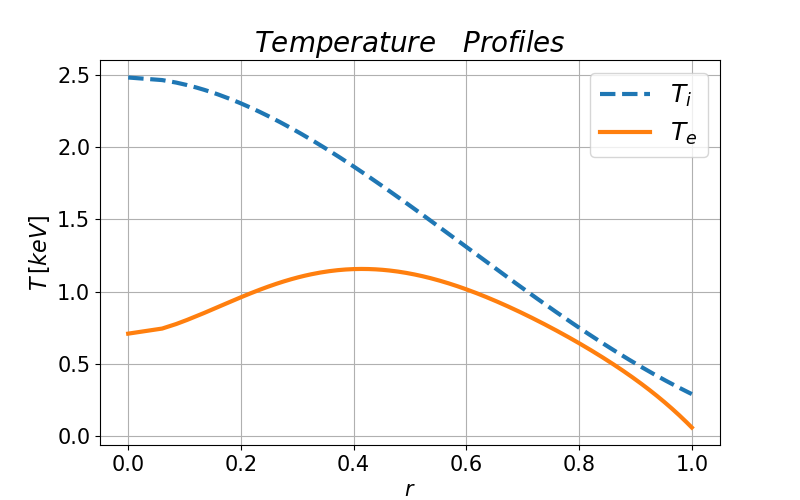}
{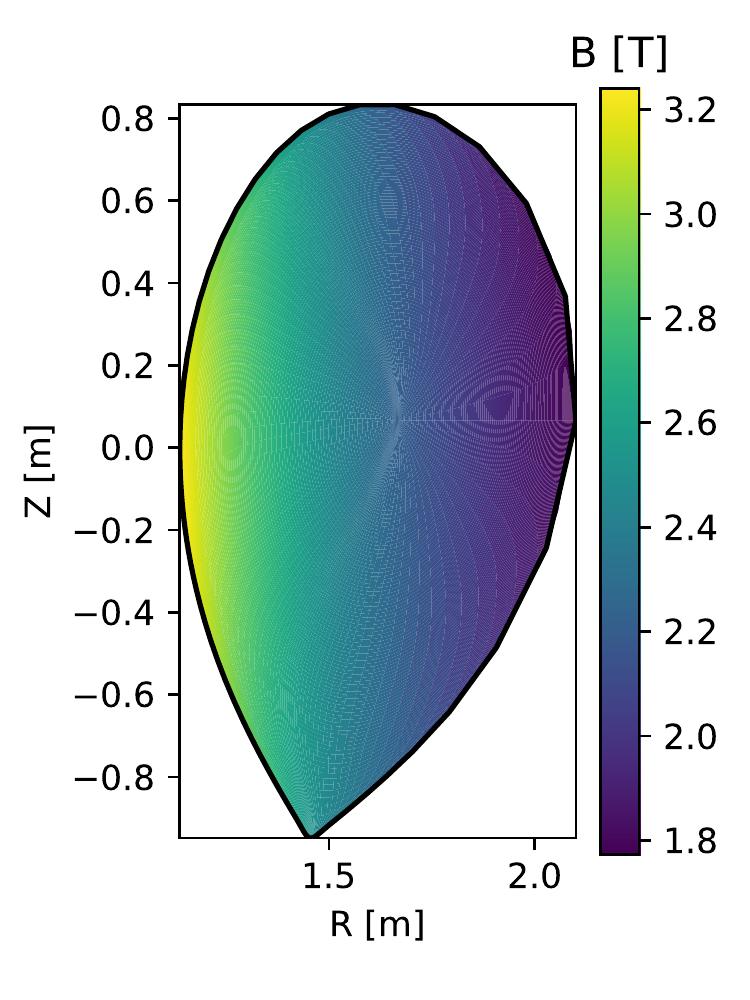}
{Top Left: Safety factor profile. Top Right: Bulk species temperature profiles. Bottom left: Poloidal view of the magnetic equilibrium in use. }
{Fig:NLED_AUG_magnetic_equilibrium}

\clearpage

\subsection{Linear simulations}
In the present subsection the results of numerical simulations  involving only the wave linear dynamics will be presented. Fig.\ref{Fig:nled_aug_case_EPM} and Fig.\ref{Fig:nled_aug_case_TAE} show the frequency spectra, mode structure and poloidal view of the scalar potential $\phi$, obtained considering respectively off-axis and on-axis density profile for the EPs and a concentration equal to $3\%$. The frequency spectra have been analyzed in the same temporal domain, when a clearly growing mode is observed. In both Fig.\ref{Fig:nled_aug_case_EPM} and Fig.\ref{Fig:nled_aug_case_TAE}  the continuum spectra obtained with the linear gyrokinetic code LIGKA \cite{[LIGKA]} is shown (red crosses), together with the analytical curve for the continuum spectra calculated in cylindrical coordinates and including the toroidicity effects, \cite{[Fu_1989]} (green dotted line). 

\begin{figure}[h!]
	\begin{center}
		\vskip -0.2em
		\includegraphics[width=0.6\textwidth]{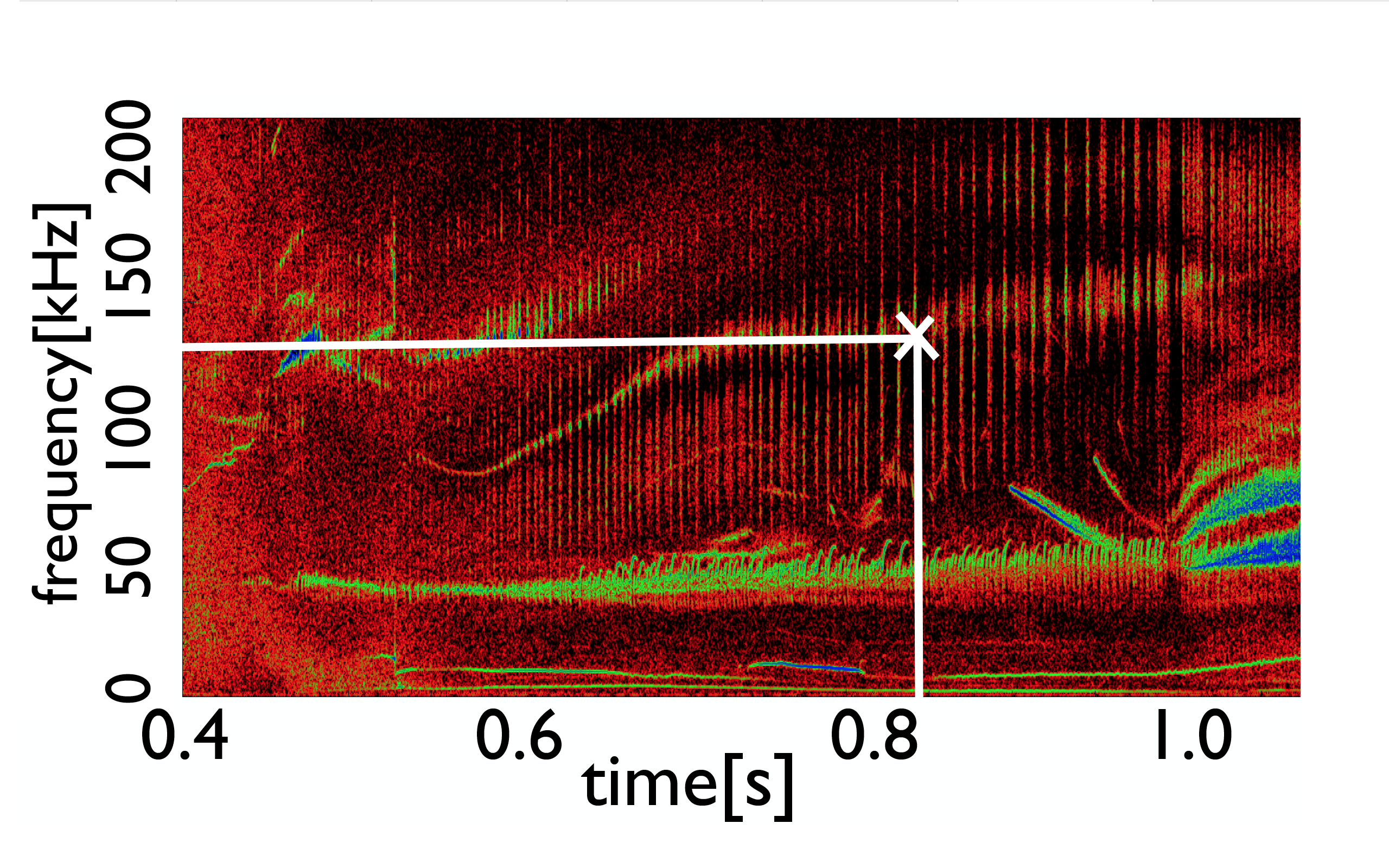}
		\vskip -1em
		\caption{Experimental spectrogram obtained with Mirnov Coil compared with theoretical prediction at one selected time. The theoretical prediction is obtained treating the fast ions as gyrokinetic.} 
		\label{Fig:spectrogram_NLED_AUG}
	\end{center}
\end{figure}

When the EPs possess an off-axis profile, a mode sitting at the radial position $r\simeq0.22$ is observed. The dominant poloidal component of the scalar potential appears to be that having $m=2$. Due to the measured frequency lying on the branch of the continuum, this mode is identified as an EPM. The frequency is measured as: 
\begin{equation}
    f = 129\,kHz
\end{equation}
When finite Larmor radius are take into account, a slightly change in the frequency (from $129$ to $131\,kHz$) is observed. The frequency measured in numerical simulations can be compared with the experimental measurements. In Fig.\ref{Fig:spectrogram_NLED_AUG}, the spectrogram obtained with Mirnov coils is shown. A big variety of EPs driven modes can be found. At $t=0.84\,s$, the modes with frequencies around $50\,kHz$ have been identified as EGAMs (see Ref.\cite{[DiSiena_2018],[Novikau_2019],[Lauber_2014]}). We focus here on the Alfv\'en modes with frequency lying in the domain between $100$ and $150\,kHz$. The numerical result shows that these modes are indeed EPMs (see the white cross in Fig.\ref{Fig:spectrogram_NLED_AUG}). Despite the approximation of the EPs distribution function we notice that the results of the numerical simulations appear to be in good agreement with the modes observed in the spectrogram.

\yFigFourMine
{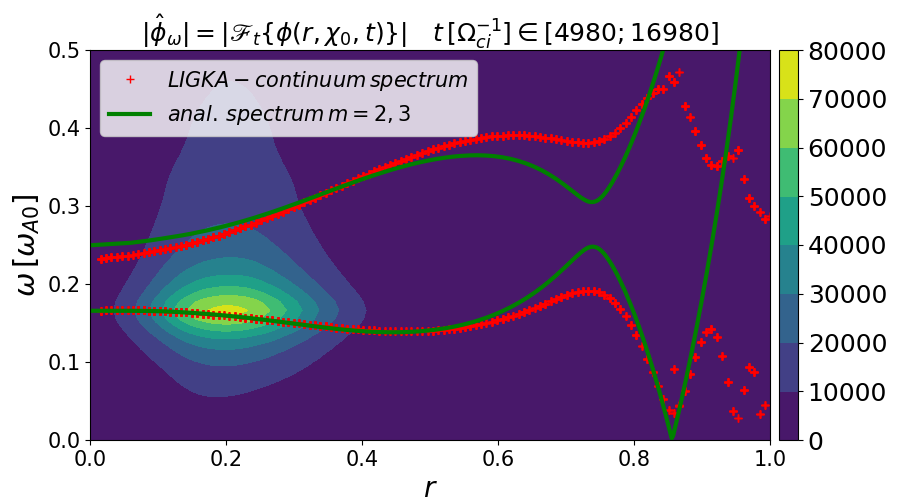}
{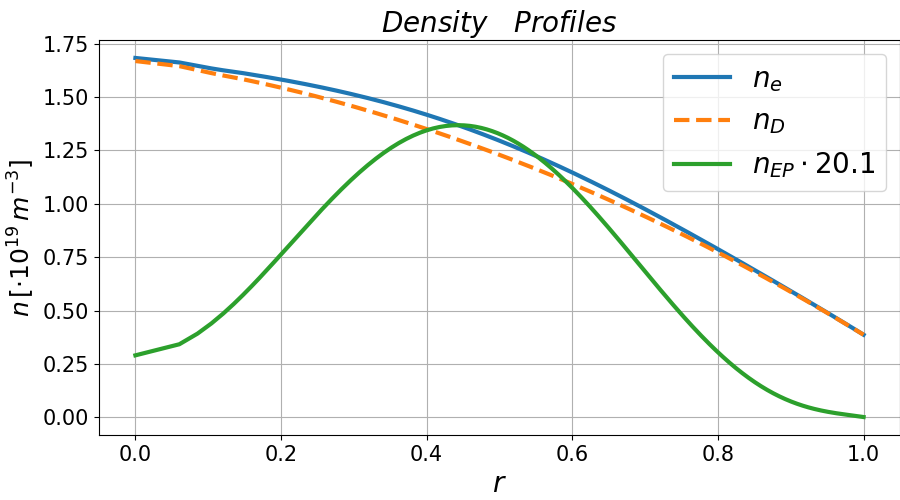}
{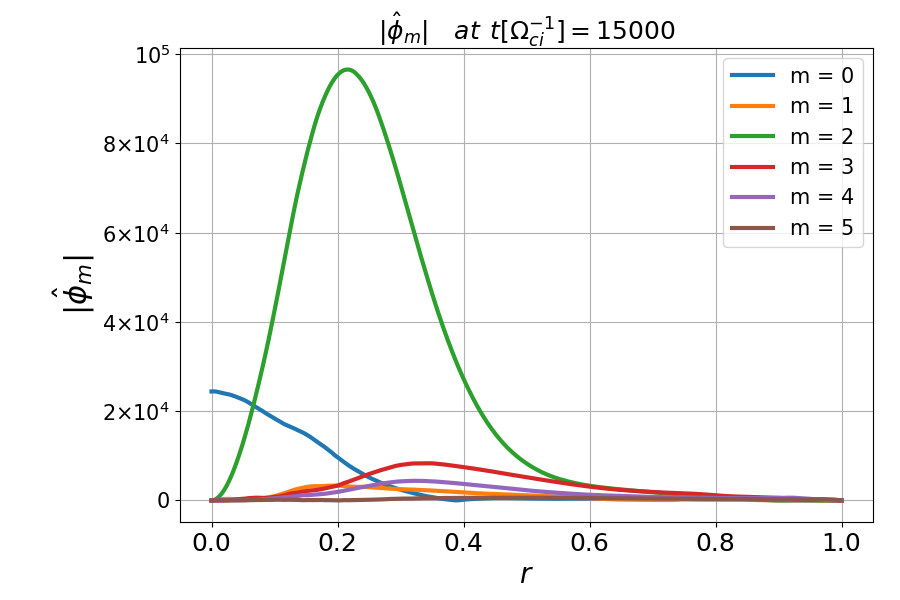}
{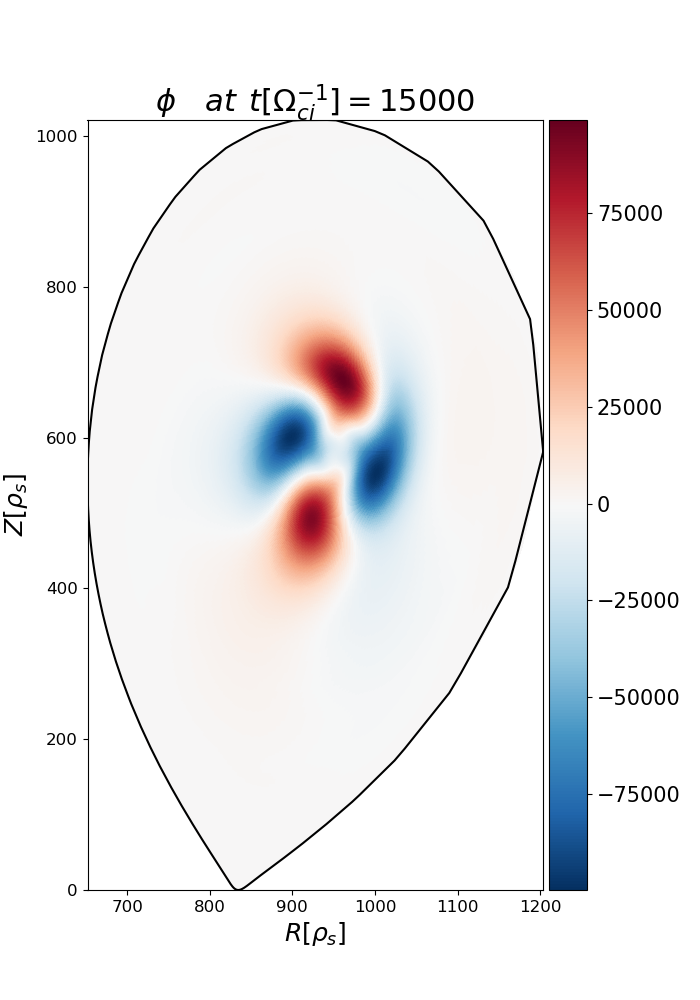}
{Numerical results for off-axis EPs profile. EPs concentration of $3\%$, $T_{EP}=113\,KeV$.}
{Fig:nled_aug_case_EPM}

\clearpage
\yFigFourMine
{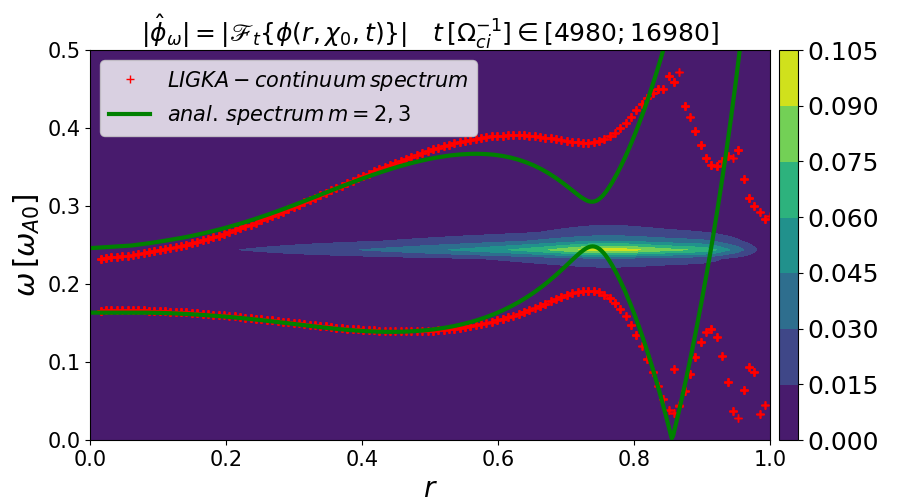}
{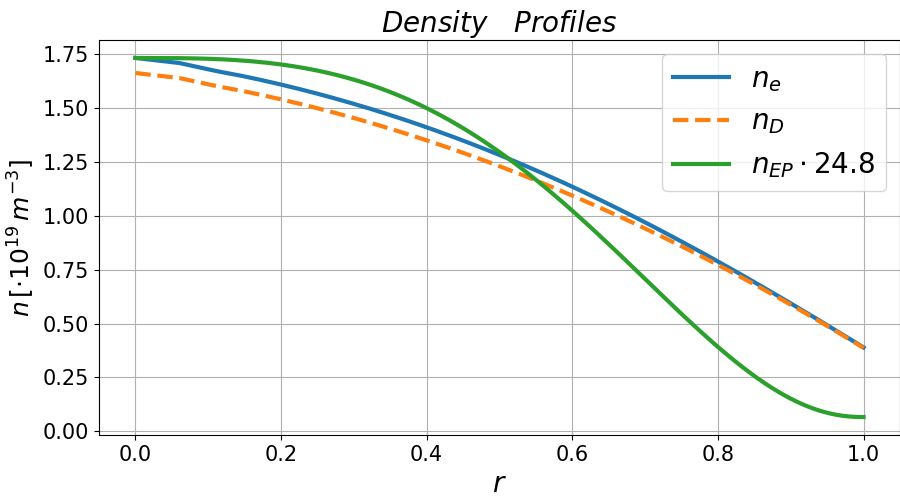}
{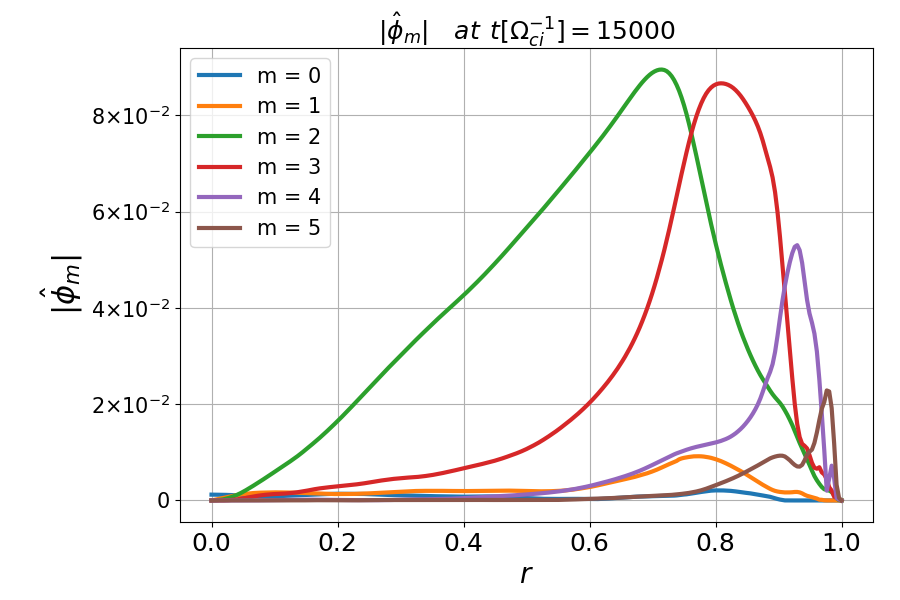}
{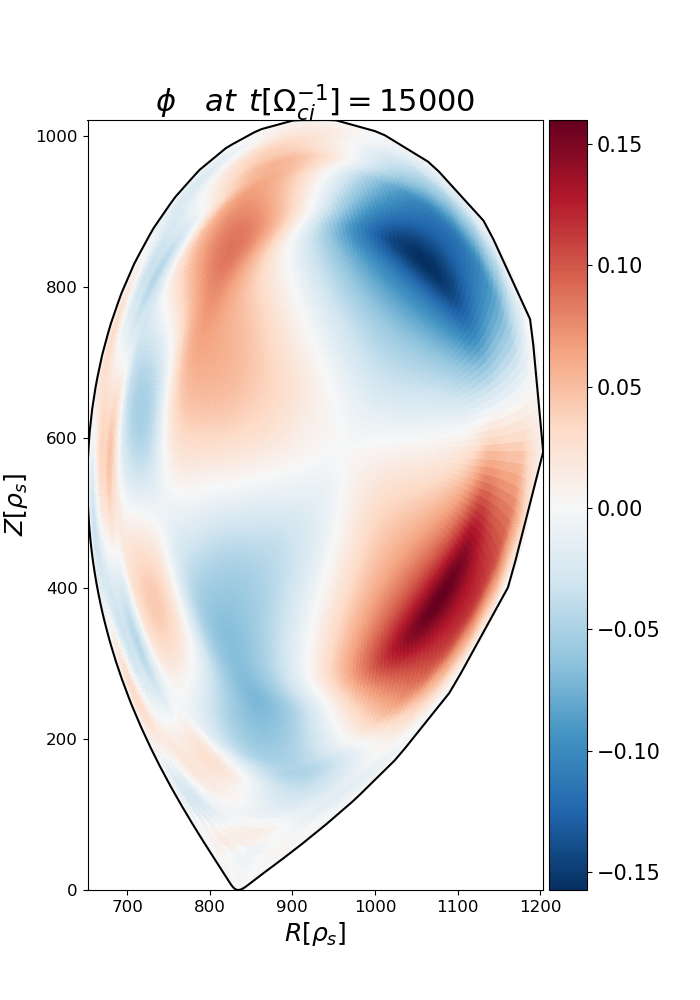}
{Numerical results for on-axis EPs profile with a concentration of $3\%$, $T_{EP}=113\,KeV$}
{Fig:nled_aug_case_TAE}

An on-axis density profile is also considered for the EPs. The radial dependence of the EPs density profiles is expressed by the formula: $n_{EP}\simeq (1-r^{\alpha})^{\beta}$. The coefficients $\alpha,\beta$ have been chosen in order to have the second derivative of $n_{EP}$ equal to zero at the position where an Alfv\'en mode is expected. The numerical analysis shows that a mode lying in the gap  of the continuum spectra, created by the poloidal modes $m=2$ and $m=3$ is observed. It appears to be peaked at the radial position $r\simeq 0.738$. Due to the radial localization and frequency this is identified as a TAE. 

In Fig.\ref{Fig:nled_aug_gamma_Te} the dependence of $\gamma$ against the value of the bulk species temperature is shown. In the plot on the top left, the dependence of the growth rate  against the electron temperature, keeping the bulk ions temperature constant ($T_{i}(r=0)=3.5\,keV$), is shown. In the plot on the top right, the dependence of the growth rate  against the bulk ion temperature, keeping the electron temperature constant ($T_{e}(r=0)=0.707\,keV$), is shown. The EPs temperature is flat $T_{EP}=113\,KeV$ and the EPs have a concentration  of $3\%$. In the plot on the bottom instead, the dependence of the damping rate (simulations without fast particles) is shown, against the value of the electron temperature. This study of the dependence of the growth or damping rate against the electron temperature shows that the electrons are the main responsible of the damping of the Alfv\'en modes even in this realistic scenario, which is identified here as electron Landau damping. Since  a realistic scenario is considered here, the approximate theoretical predictions for the Landau damping described in the previous sections is outside its validity regime and therefore is not shown.

\yFigThree
{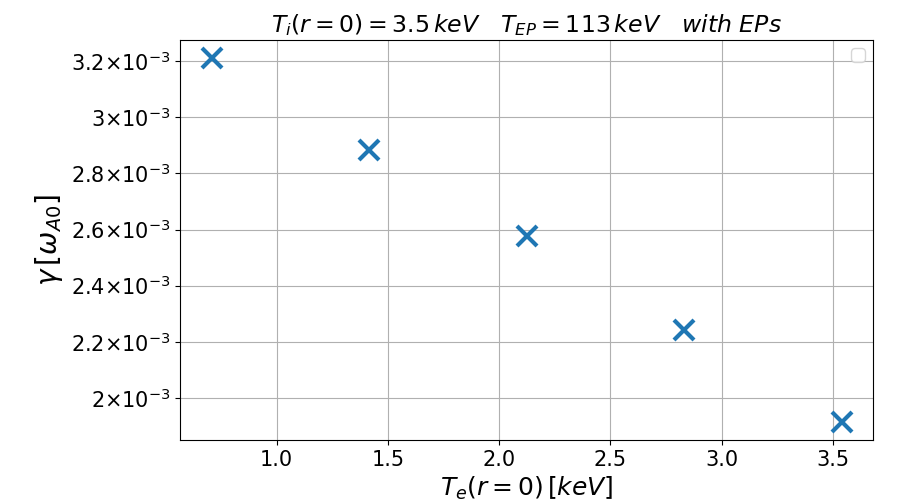}
{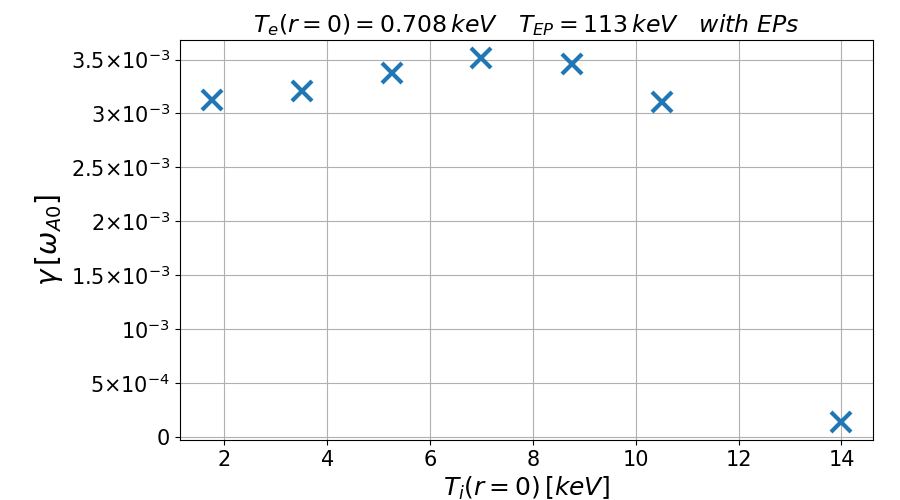}
{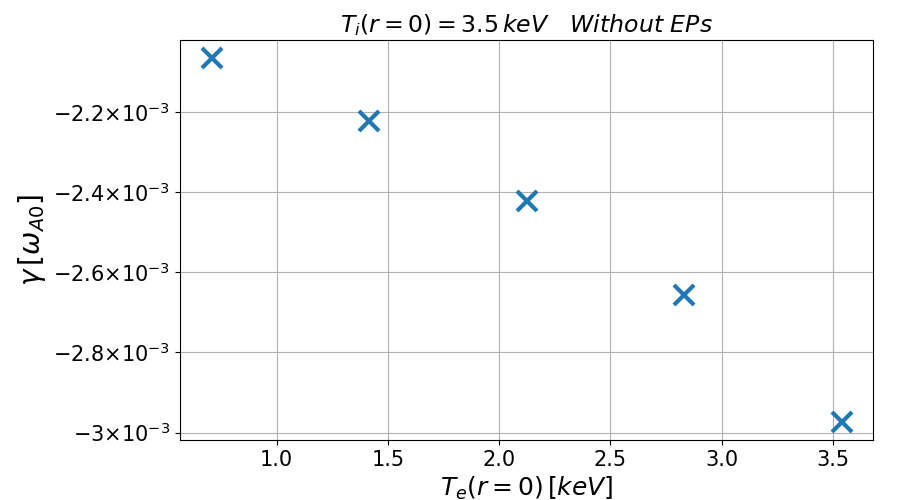}
{Top: respectively on the left and on the right, two scans in the electron and bulk ions temperature for a growing mode are shown. EPs concentration  $3\%$,$T_{EP}=113\,KeV$. Bottom: Scan in the electron temperature for a damped mode.}
{Fig:nled_aug_gamma_Te}


In Fig.\ref{Fig:continuum_vs_landau_nled} a theoretical estimation of the regions in the radial domain where the Landau damping is supposed to dominate over the continuum damping is presented for simulations without EPs. In order to perform this calculation, the characteristic radial structure has been measured in a simulation with an EPM, giving the value of $k_{r,0}$.  This allows us to calculate the analytical prediction for the half decay time of the continuum damping. The half decay time due to Landau damping, on the other hands, can be measured in a simulation without EPs:
\begin{align}
    t_{1/2,continuum\,damping}=\frac{\abs{2-k_{r,0}}}{\abs{\frac{\partial\omega}{\partial s}}} && t_{1/2,Landau\,damping}=\log{2}/\gamma
\end{align}
The radial regions where the Landau damping dominates over the continuum damping are those where the half decay time of the Landau damping is smaller than the half decay time of the continuum damping:  $t_{1/2,continuum\,damping}\,\textgreater\,t_{1/2,Landau\,damping} $. Equivalently it can be said that the Landau damping dominates over the continumm damping in those regions where:
\begin{equation}
    \gamma\,\textgreater\,\abs{\frac{\log(2)}{2-k_{r,0}}}\cdot \abs{\frac{\partial\omega}{\partial s}}
    \label{Eq:nled_gamma_greater}
\end{equation} 
In Fig.\ref{Fig:continuum_vs_landau_nled} on the left the continuum spectra calculated with the code LIGKA are shown. They are used to calculate the derivative of the frequency to obtain the estimation of the regions where the Landau damping dominates over the continuum damping. To do so, the frequency has been divided in two branches, denoted as upper and lower branch. In Fig.\ref{Fig:continuum_vs_landau_nled} on the right the regions where $t_{1/2,continuum\,damping}\,\textgreater\,t_{1/2,Landau\,damping} $ are shown. Therefore the green regions correspond to the radial domain where  the Landau damping is dominant over the continuum damping if the frequency of the mode  is sitting on the lower branch. The cyan regions instead represent the domains where the Landau damping is dominating on the continuum damping if the frequency of the mode is sitting on the upper branch.

\yFigTwo
{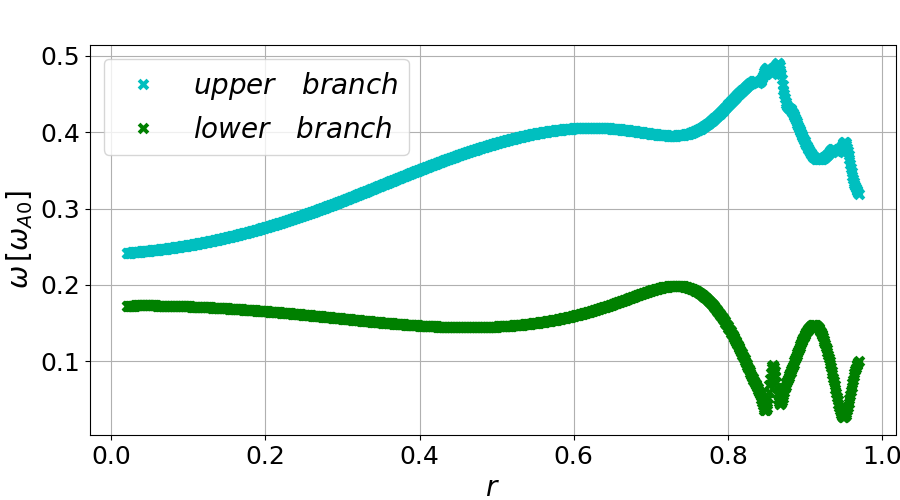}
{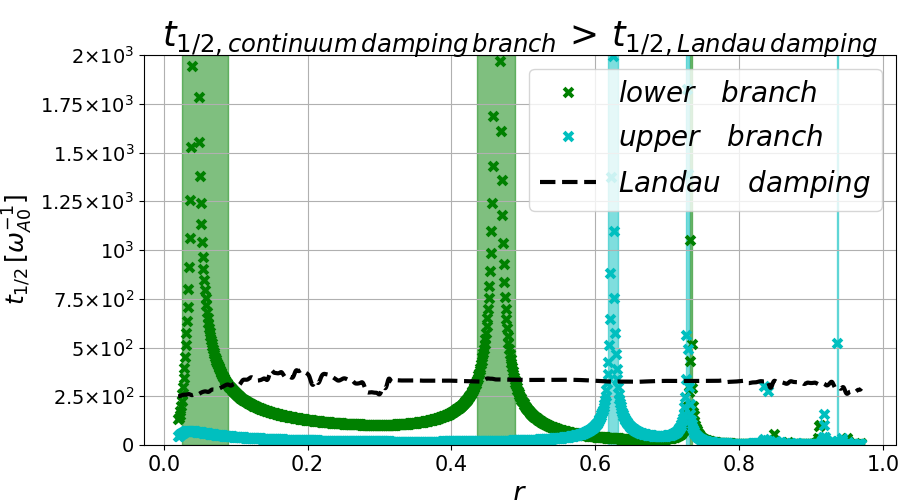}
{Left: continuum spectra calculated with the code LIGKA. It has been divided into two branches (denoted as upper and down). Right: regions where $t_{1/2,continuum\,damping}\,\textgreater\,t_{1/2,Landau\,damping} $.}
{Fig:continuum_vs_landau_nled}

Finally in Fig.\ref{Fig:Gamma_Ligka_Orb}, the results of a first benchmark between ORB5 and the code LIGKA are shown. Here a scan in the EPs temperature is depicted. The EPs have an on-axis profiles and their concentration is kept constant and equal to $3\%$. A reasonable agreement has been found between the two codes for the measured growth rate and frequency.

\yFigTwo
{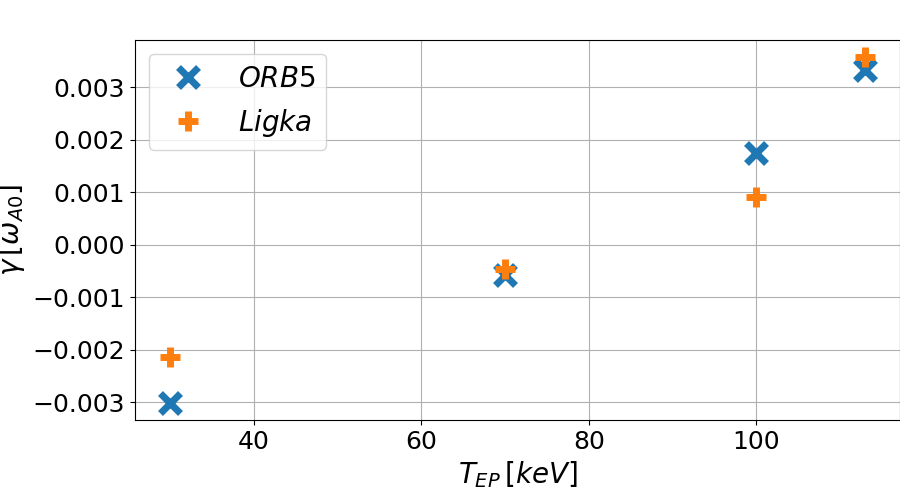}
{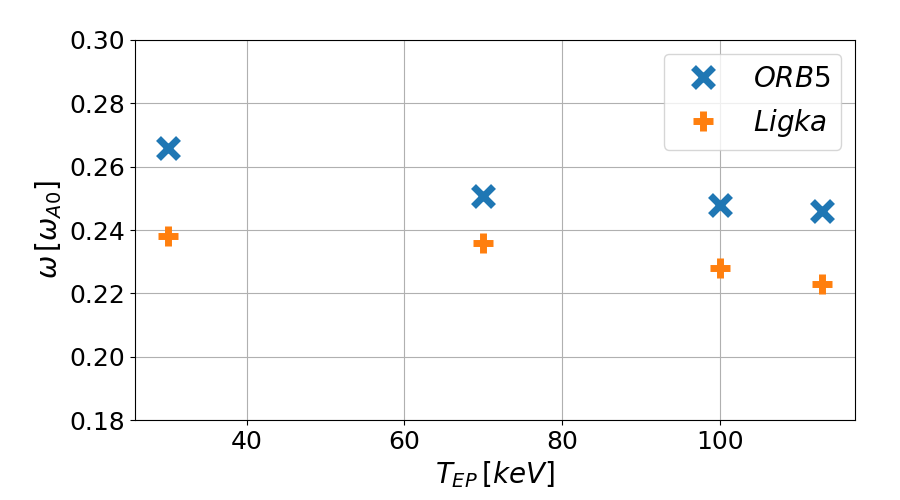}
{Scan in $T_{EP}$. TAE growth rate and frequency, calculated with LIGKA and ORB5 for an EPs concentration equal to $3\,\%$ (same density, temperature profiles in use).}
{Fig:Gamma_Ligka_Orb}

\clearpage
\subsection{Nonlinear simulations}
In this subsection results involving the nonlinear dynamics of the Alfv\'en waves are presented, when both on-axis (Fig.\ref{Fig:TAE_nl}) and off-axis (Fig.\ref{Fig:EPM_nl}) density profiles for the EPs are considered. With an on-axis density profiles of the EPs, a mode sitting in the frequency gap is observed (TAE),  Fig.\ref{Fig:TAE_nl}. Its mode structure and frequency spectra are not observed to change passing from the linear to the nonlinear phase, confirming its nature of an eignemode of this system, which is only weakly perturbed by EPs.
\yFigFour
{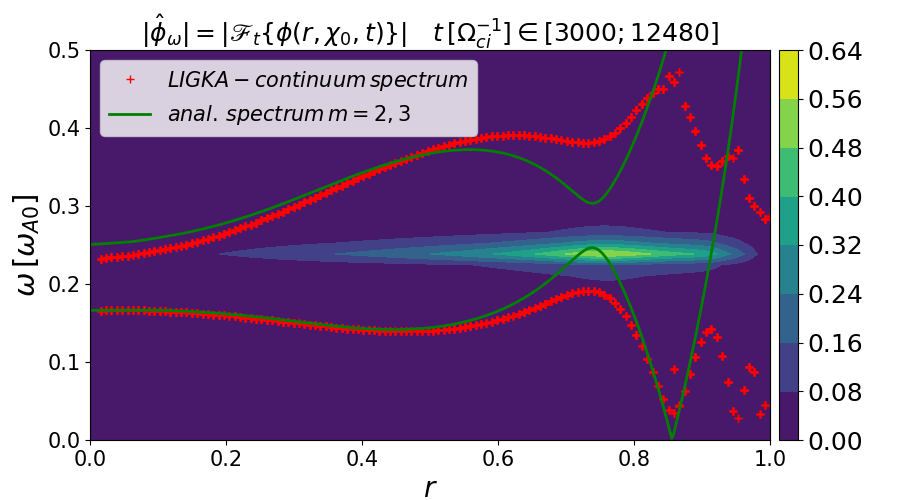}
{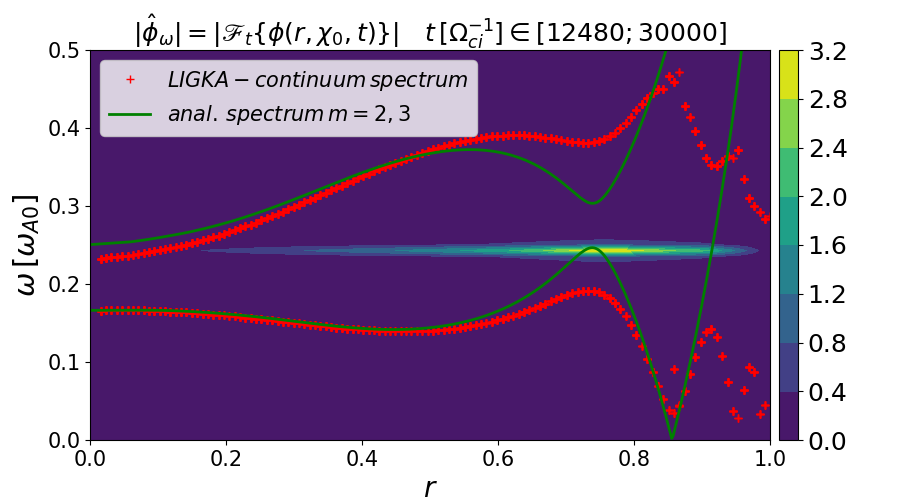}
{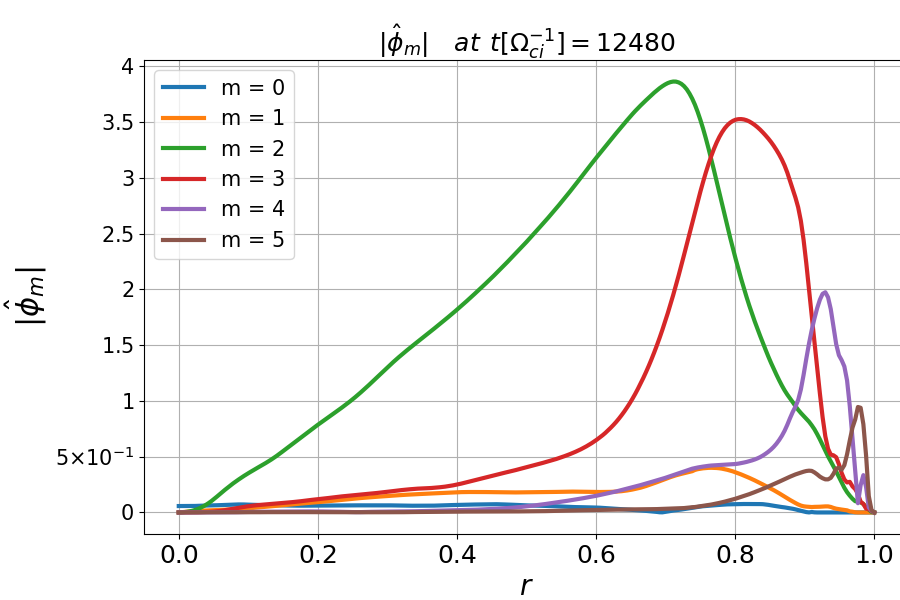}
{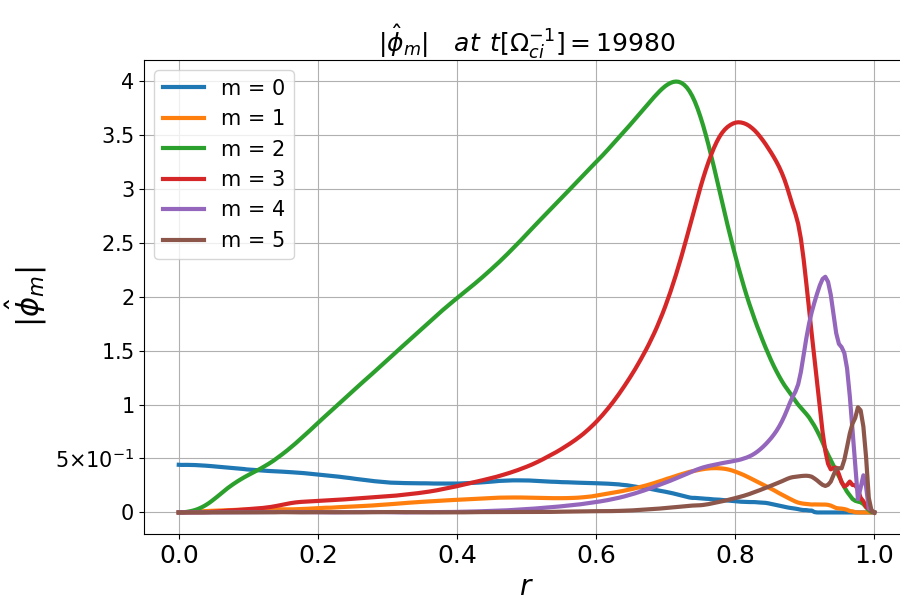}
{Mode structure and frequency spectra in the linear phase (left) and in the saturation phase (right). EPs have ah on-axis profile.}
{Fig:TAE_nl}

\begin{figure}[h!]
	\begin{center}
		\vskip -0.2em
		\includegraphics[width=0.6\textwidth]{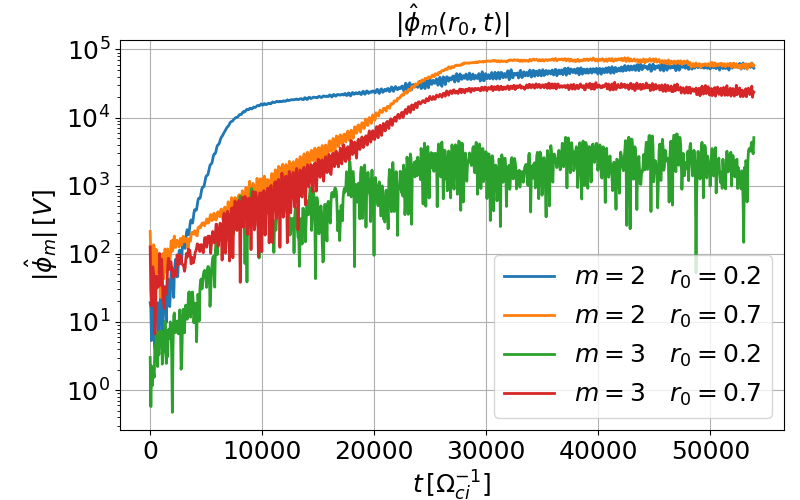}
		\vskip -1em
		\caption{Time evolution of the dominant poloidal modes of the scalar potential ($m=2,3$) at the radial positions where the TAE and EPM are located. EPs have an off-axis profiles. }
		\label{Fig:EPM_nl_time_evolution}
	\end{center}
\end{figure}

When an off-axis density profile for the EPs is considered, a mode with dominant poloidal mode number $m=2$ and peaked around $r\simeq 0.22$ is observed (see Fig.\ref{Fig:EPM_nl}). This is consistent with what was observed in the previous sections, when just the linear effects in the simulations were involved. Passing to the nonlinear phase a secondary mode with $m=2$ and $m=3$ is observed to grow around the radial position $r\simeq 0.738$. This second mode is identified as the previously described TAE. This happens, because in the first linear phase the EPs drive  the EPM unstable, which appears in fact to be dominant. In the nonlinear phase, the coexistence of the EPM and TAE is observed, due to an earlier saturation of the EPM (see Fig.\ref{Fig:EPM_nl_time_evolution}) .

\yFigFour
{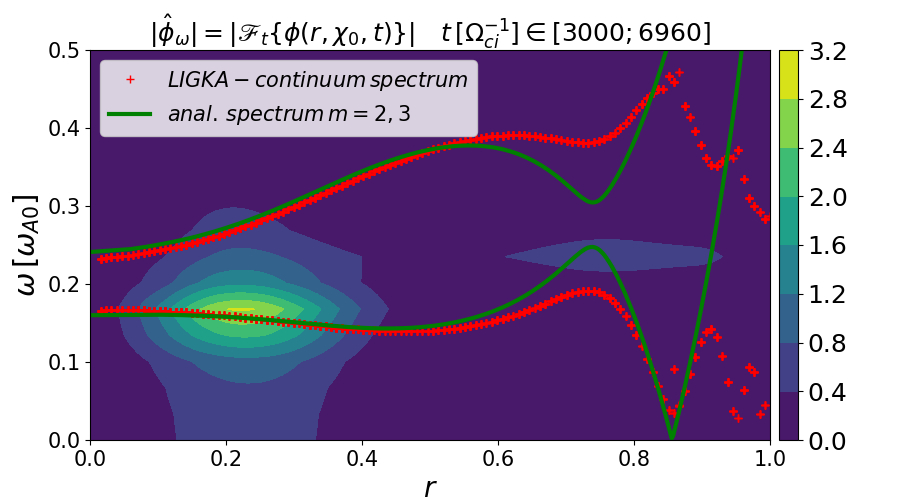}
{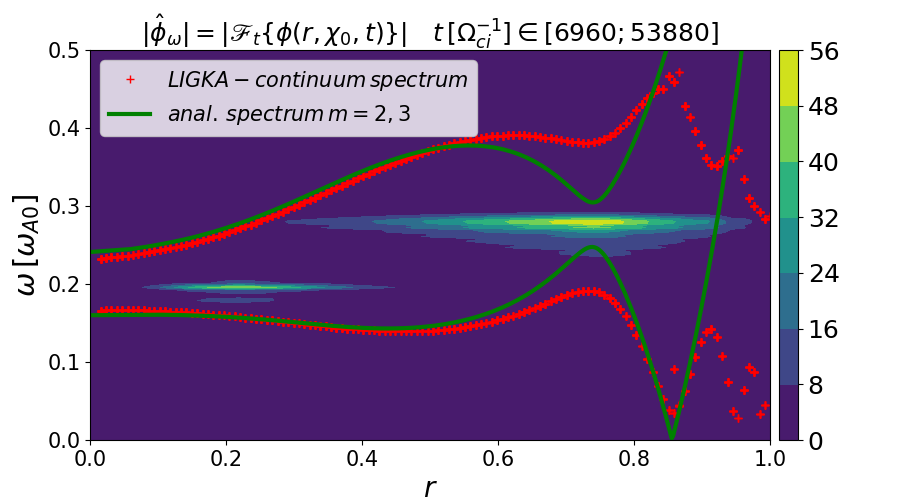}
{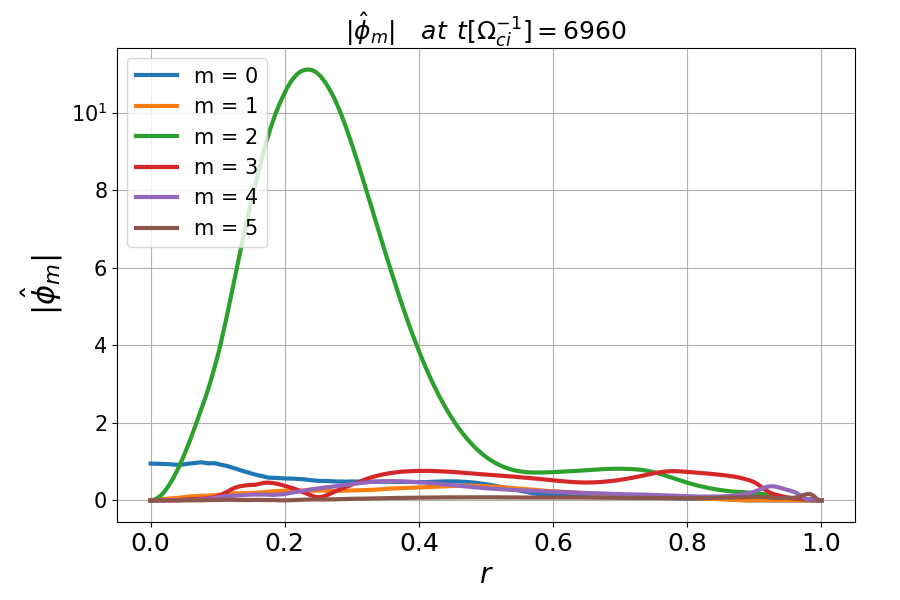}
{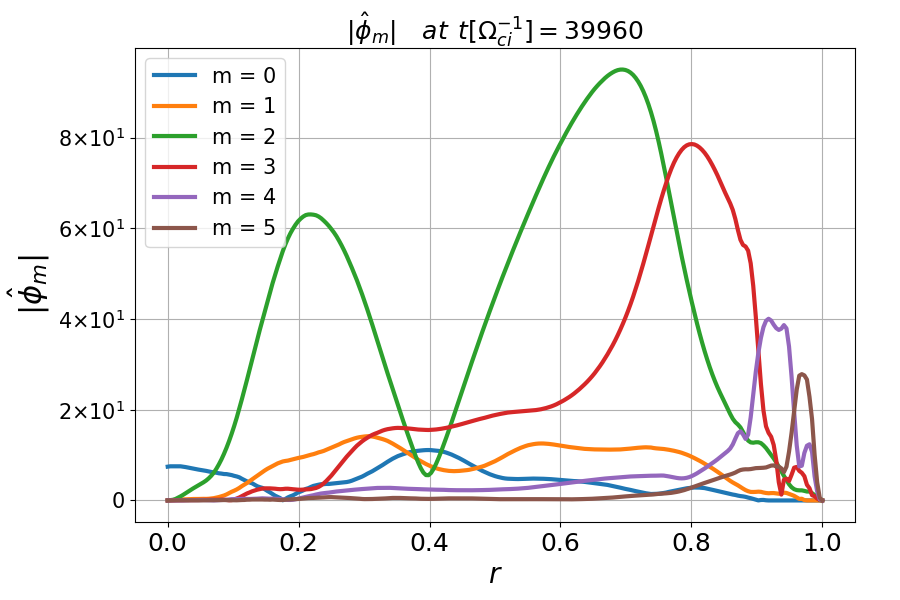}
{Mode structure and frequency spectra in the linear phase (left) and in the saturation phase (right). EPs have an off-axis profiles.}
{Fig:EPM_nl}

\clearpage
\section{Conclusion}
The presence of Alfv\'en modes in burning plasma can affect negatively the energy confinement and can also cause a damage in the confining machine. Because of their importance, the present paper has dealt with the main damping mechanisms affecting the Alfv\'en modes, trying to outline them and to understand in which regime they are acting. Among the great zoology of existing Alfv\'en modes, the attention has been focused on toroidal Alfv\'en eigenmodes and energetic particle modes.  These studies have been mainly carried by means of numerical simulations conducted with the code ORB5. The obtained results have been compared, when possible, with the presented analytic theory developed in a simplified geometry, and with the results of the linear code Ligka.

In Section 3 simulations with very small inverse aspect ratio ($\epsilon=0.01$) have been considered, in order to lead the analysis in the cylinder limit. Simplified profiles have been taken into account and very low electron temperature has been considered in order to have the continuum damping dominant over the Landau damping. The developed theory has been used to analyze the results of simulations without energetic particles. The dependence of the radial wave number of the mode against the time have been observed (phase mixing). Also the scalar potential has been found to decay as $
\delta\phi\propto \abs{\omega_{A}^{\prime}\,t}^{-1}$ (continuum damping).

In Section 4 higher bulk ion and electron temperatures have been considered, in order to observe the Landau damping to be dominant over the continuum damping. The numerical simulations have been conducted using plasma equilibrium and profiles from the ITPA-TAE international benchmark case. In order to separate the ions and electrons contributions to the damping, a kinetic term to the ideal MHD vorticity equation has been added. Following a perturbative approach, a simplified estimation for the Landau damping has been obtained using cylindrical coordinates. The developed analytical theory has been compared with the dependence found in numerical simulations of the damping rate against the bulk electron temperature. A reasonable agreement has been found and and this has also proved that the electron are the main responsible for the damping. 

In Section 5 a realistic plasma equilibrium taken from a shot in ASDEX Upgrade has been considered. The results of the linear numerical simulations have shown the dependence of the damping rate against the bulk electron temperature describing, also in this case, the action of the Landau damping. A theoretical estimation of the radial regions where the Landau damping is expected to be dominant over the continuum damping has been presented, for simulations near marginal stability. A benchmark with the code LIGKA has shown good agreement for the frequency and growth rate dependence on the EPs temperature. Finally, the nonlinear simulations have shown the interaction of an EPM and a TAE in the scenario with off-axis EPs density profile. 

The future works will extend the developed theory in order to find a better agreement between the predicted estimation of the decaying mode and the numerical simulations. In Ref.\cite{[Chen-16]} it was suggested that all the Alfv\'en fluctuations can be explained within the framework of a single general fishbone-like dispersion relation (GFLDR).  This could represent the starting point  to improve the anaytical prediction of the damping rate and it would be a very interesting analytical and numerical task. Future and deeper benchmark with the code Ligka and the Hybrid Magnetohydrodynamics Gyrokinetic code HYMAGYC \cite{[HYMAGYC]} would be of great interest in order to better understand the linear and nonlinear dynamics contained in the NLED-AUG case.

\section{Acknowledgments}
Simulations presented in this work were performed on the CINECA Marconi supercomputer within the ORBFAST and OrbZONE projects. 

One of the authors, F. Vannini, would like to thank  Xin Wang, Zhixin Lu   and Gregorio Vlad for useful, interesting discussions and for great help provided in understanding Alfv\'en dynamics, Gyrokinetic and MHD theory. The authors wish to acknowledge stimulating discussions with F. Zonca, G. Fogaccia, A. K\"onies, J. Gonzalez-Martin and A. Di Siena. This work was partly performed in the frame of the  \quotes{Multi-scale Energetic particle Transport in fusion devices} ER project.

This work has been carried out within the framework of the EUROfusion Consortium and has received funding from the Euratom research and training program 2014-2018 and 2019-2020 under grant agreement number 633053. The views and opinions expressed herein do not necessarily reflect those of the European Commission. 
\clearpage


\clearpage
\begin{thebibliography}{8}

\bibitem{[Bierwage-2018]} A. Bierwage, Kouji Shinhoara, Yasushi Todo, Nobuyuki Aiba, Masao Ishikawa, Go Matsunaga, Manabu Takechi and Masatoshi Yagi, \quotes{Simulation tackle abrupt massive migrations of energetic beam ions in tokamak plasma}, {\it Nature Communications} {\bf 9}, (2018) 

  	
\bibitem{[Chen-16]} Liu Chen and Fulvio Zonca, \quotes{Physics of Alfv\'en waves and energetic particles in burning plasmas}, {\it Rev. Mod. Phys.} {\bf 88}, (2016)
  	
\bibitem{[Jolliet-07]} S. Jolliet, A. Bottino, P. Angelino, R. Hatzky, T.M. Tran, B. McMillan, O. Sauter, K. Appert, Y. Idomura and L. Villard, \quotes{A global collisionless PIC code in magnetic coordinates}, {\it Comput. Phys. Comm.} {\bf 177}, (2007)
  	
\bibitem{[Bottino-11]} A. Bottino, T. Vernay, B. Scott, S. Brunner, R. Hatzky, S. Jolliet, B.F. McMillan, T.M. Tran and L. Villard, \quotes{Global simulations of tokamak microturbulence: finite-$\beta$ effects and collisions}, {\it Plasma Phys. Control. Fusion} {\bf 53}, (2011)
  	
  	
\bibitem{[LIGKA]} P. Lauber, S. G\"unter, A. K\"onies and S. D. Pinches, \quotes{LIGKA: A linear gyrokinetic code for the description of background kinetic and fast particle effects on the MHD stability in tokamaks} {\it Journal of Computational Physics} {\bf 226}, (2007)
  	
\bibitem{[Koenies_2018]} A. K\"onies, S. Briguglio, N. Gorelenkov, T. Fehér, M. Isaev, Ph. Lauber, A. Mishchenko, D.A. Spong, Y. Todo, W. A. Cooper, R. Hatzky, R. Kleiber, M. Borchardt, G. Vlad, A. Biancalani, A. Bottino and ITPA EP TG, \quotes{Benchmark of gyrokinetic, kinetic MHD and gyrofluid codes for the linear calculation of fast particle driven TAE dynamics} {\it Nucl. Fusion} {\bf 58}, (2018)
  	
  	
\bibitem{[Lauber]} Ph. Lauber,http:\url{//www2.ipp.mpg.de/~pwl/NLED_AUG/data.html}
  	
  		
\bibitem{[Lanti-2019]} E. Lanti, N. Ohana, N. Tronko, T. Hayward-Schneider, A. Bottino, B.F. McMillan, A. Mishchenko, A. Scheinberg, A. Biancalani, P. Angelino, S. Brunner, J. Dominski, P. Donnel, C. Gheller, R. Hatzky, A. Jocksch, S. Jolliet, Z. Lu, J Collar and L. Villard, \quotes{ORB5: a global electromagnetic gyrokinetic code using the PIC approach in toroidal geometry}, {\it Computer Physics Communications}, (2019) 
  	
\bibitem{[Tronko-2016]} N. Tronko, A. Bottino and E. Sonnendr\"{u}cker, \quotes{Second order gyrokinetic theory for particle-in-cell codes}, {\it Phys. Plasmas} {\bf 23}, (2016)
  	
\bibitem{[Myato-2013]} N. Myato, B. Scott and Masatoshi Yagi, \quotes{On the gyrokinetic model in long wavelength regime}, {\it Plasma Physics and Controlled Fusion} {\bf 55}, (2013)
  
\bibitem{[CHEASE]} H. L\"{u}tjens, A. Bondeson and O. Sauter, \quotes{The CHEASE code for toroidal MHD equilibria},{\it Comput. Phys. Commun.} {\bf 97}, (1996)
  	
\bibitem{[Bottino-2015]} A. Bottino and E. Sonnendr\"{u}cker, \quotes{Monte Carlo particle-in-cell methods for the simulation of Vlasov-Maxwell gyrokinetic equations}, {\it J. Plasma Phys.} {\bf 81}, (2015)
  

\bibitem{[Tronko-2017]} N. Tronko, A. Bottino, C. Chandre and E. Sonnendr\"{u}cker, \quotes{Hierarchy of second order gyrokinetic Hamiltonian models for particle-in-cell codes}, {\it Plasma Phys. Control. Fusion} {\bf 59}, (2017)

  	
\bibitem{[Mishchenko-PB]} A. Mishchenko, A. Bottino, A. Biancalani,  R. Hatzky, T. Hayward-Schneider, N. Ohana, E. Lanti, S. Brunner, L. Villard, M. Borchardt et al., \quotes{Pullback scheme implementation in ORB5}, {\it Comput. Phys. Commun.} {\bf 238}, (2019)
  	
\bibitem{[Zonca_Vlad_99]} G. Vlad, F. Zonca and S. Briguglio, \quotes{Dynamics of Alfv\'en waves in tokamaks}, {\it La Rivista del Nuovo Cimento} {\bf 22}, (2008)
  	
\bibitem{[Zonca_2008]} Fulvio Zonca and Liu Chen, \quotes{Radial structures and nonlinear excitation of geodesic acoustic modes}, {\it Europhysics Letters} {\bf 83}, (2008)
  	
\bibitem{[Palermo_2016]} F. Palermo, A. Biancalani,  C. Angioni, F. Zonca and A. Bottino, \quotes{Combined action of phase-mixing and Landau damping causing strong decay of geodesic acoustic modes}, {\it Europhys. Lett. } {\bf 115},  (2016)
  	
\bibitem{[Biancalani_2016]} A. Biancalani, F. Palermo, C. Angioni, A. Bottino and F. Zonca, \quotes{Decay of geodesic acoustic modes due to the combined action of phase mixing and Landau damping}, {\it Phys. Plasmas } {\bf 23}, (2016)
  	
\bibitem{[Chen_1974]} Liu Chen  and Akira Hasegawa, \quotes{Plasma heating by spatial resonance of Alfv\'en wave}, {\it Physics of Fluids} {\bf 17}, (1974)
  	
\bibitem{[Fu_1989]} G. Y. Fu  and J. W. Van Dam, \quotes{Excitation of the toroidicity-induced shear Alfv\'en eigenmode by fusion alpha particles in an ignited tokamak}, {\it Physics of Fluids B: Plasma Physics} {\bf 10}, (1989)
  
  	
\bibitem{[Koles_2002]} Ya. I. Kolesnichenko, V. V. Lutsenko, H. Wobig and V. Yakovenko, \quotes{Alfv\'en instabilities driven by circulating ions in optimized stellarators and their possible consequences in a Helias reactor} {\it Physics of Plasmas} {\bf 9}, (2002)
  	
\bibitem{[Betti_1992]} R. Betti and J. P. Freidberg, \quotes{Stability of Alfv\'en gap modes in burinig plasmas} {\it Physics of Fluids B: Plasma Physics} {\bf 4}, (1992)
  	
\bibitem{[Koles_2016]} Ya. I. Kolesnichenko, A. K\"{o}nies, V. V. Lutsenko, M. Drevlak, Yu. Turkin and P. Helander, \quotes{Isomon instabilities driven by energetic ions in Wendelstein 7-X} {\it Nucl. Fusion} {\bf 56}, (2016)
  	
 
  
 
\bibitem{[DiSiena_2018]} A. Di Siena, A. Biancalani, T. G\"orler, H. Doerk, I. Novikau, P. Lauber, A. Bottino and E. Poli, \quotes{Effect of elongation on energetic particle-induced geodesic acoustic mode} {\it Nucl. Fusion} {\bf 58}, (2018)

\bibitem{[Novikau_2019]} I. Novikau, A. Biancalani, A. Bottino, A. Di Siena, P. Lauber, E. Poli, E. Lanti, L. Villard, N. Ohana and S. Briguglio, \quotes{Implementation of energy transfer technique in ORB5
to study collisionless wave-particle interactions in phase-space} {\it Comp. Phys. Comm.} ,  (2019) submitted


\bibitem{[Lauber_2014]} P. Lauber, \quotes{Off-axis NBI heated discharges at ASDEX Upgrade: EGAMs, 
RSAEs, TAE bursts} {\it 13th Energetic Particle Physics TG Meeting, Padua, 
Italy, 21–23 October}, (2014)

\bibitem{[HYMAGYC]} G. Fogaccia, G. Vlad and S. Briguglio, \quotes{Linear benchmarks between the hybrid codes HYMAGYC and HMGC to study energetic particle driven Alfv\'enic modes} {\it Nucl. Fusion} {\bf 56}, (2016)

  
\end{thebibliography}
\end{document}